%% file: arxiv_jfm.tex
\documentclass{jfm}
\usepackage{graphicx}
\usepackage{rotating,amsmath,float,upmath}
\usepackage{natbib}
\usepackage{color}
\usepackage{tikz}
\usepackage{array}
\newcolumntype{?}{!{\vrule width 1pt}}
\usepackage{booktabs} 
\usepackage{tabu}
\usepackage{url} 
\usepackage{breqn}
\usepackage{mwe}

\ifCUPmtlplainloaded \else
  \checkfont{eurm10}
  \iffontfound
    \IfFileExists{upmath.sty}
      {\typeout{^^JFound AMS Euler Roman fonts on the system,
                   using the 'upmath' package.^^J}%
       \usepackage{upmath}}
      {\typeout{^^JFound AMS Euler Roman fonts on the system, but you
                   dont seem to have the}%
       \typeout{'upmath' package installed. JFM.cls can take advantage
                 of these fonts,^^Jif you use 'upmath' package.^^J}%
      }
  \else
  \fi
\fi

\ifCUPmtlplainloaded \else
  \checkfont{msam10}
  \iffontfound
    \IfFileExists{amssymb.sty}
      {\typeout{^^JFound AMS Symbol fonts on the system, using the
                'amssymb' package.^^J}%
       \usepackage{amssymb}%
         \let\leq=\leqslant
         \let\geq=\geqslant
      }{}
  \fi
\fi

\ifCUPmtlplainloaded \else
  \IfFileExists{amsbsy.sty}
    {\typeout{^^JFound the 'amsbsy' package on the system, using it.^^J}%
     \usepackage{amsbsy}}
    {\providecommand\boldsymbol[1]{\mbox{\boldmath $##1$}}}
\fi

\DeclareMathAlphabet\mathbfcal{OMS}{cmsy}{b}{n}

\input{symdef.inc}

\def \ii {\mathrm{i}}

\def \mL {\mathcal{L}}

\def \mbarmu {\breve{\mu}}
\def \Ep {E^{\prime}}

\def \baromg {\bar{\Omega}}
\def \baromgamp {\bar{\Omega}^{\mathrm{mag}}}

\def \barE {\bar{E}}

\def \Up {\mathbf{U}}
\def \Omp {\boldsymbol{\Omega}}
\def \bFfp {\mathbf{F}^{\mathrm{f}\rightarrow\mathrm{p}}}
\def \bTfp {\boldsymbol{\Gamma}^{\mathrm{f}\rightarrow\mathrm{p}}}
\def \Ffp {F^{\mathrm{f}\rightarrow\mathrm{p}}}
\def \Tfp {\Gamma^{\mathrm{f}\rightarrow\mathrm{p}}}

\def \nTfp {\bar{\Gamma}^{\mathrm{f}\rightarrow\mathrm{p}}}
\def \nTvis {\bar{\Gamma}^{\mathrm{hydro}}}

\def \nE {\bar{E}}
\def \nP {\bar{\mathcal{P}}}
\def \dt {\Delta t}
\def \tp {t_0}

\def \ortp {\be_\mathrm{p}}
\def \bN {\mathbf{N}}
\def \bQ {\mathbf{Q}}
\def \nont {\bar{t}}

\def \esl {\epsilon_{\mathrm{sl}}}
\def \nonbr {\bar{\br}}
\def \nonT {\bar{T}}
\def \nons {\bar{s}}
\def \bein {\beta_{\mathrm{p}}}

\def \nonbFfp {\bar{\bF}^{\mathrm{f}\rightarrow\mathrm{p}}}
\def \nonbUp {\bar{\bU}}
\def \nonbTfp {\bar{\boldsymbol{\Gamma}}^{\mathrm{f}\rightarrow\mathrm{p}}}

\def \bTel {\boldsymbol{\Gamma}^{\mathrm{elec}}}

\def \bxp {\bx_{\mathrm{p}}}
\def \nonbP{\bar{\bP}}
\def \nonbE{\bar{\bE}}

\def \nonbOmega {\bar{\boldsymbol{\Omega}}}
\def \nonOmega {\bar{\Omega}}

\def \nonbxp {\bar{\bx}_{\mathrm{p}}}
\def \be {\mathbf{e}}

\def \bstheta {\hat{\theta}}
\def \bsP {\hat{\mP}}
\def \ptheta {\theta^{\prime}}
\def \pthetaone {\theta_1^{\prime}}

\def \pthemag {\tilde{\theta}}

\def \pPQ {\mP^{\prime}_Q}
\def \pPT {\mP^{\prime}_3}

\def \Ecri {E^{\mathrm{cri}}}

\def \sFfp {\tilde{F}^{\mathrm{f}\rightarrow\mathrm{p}}}
\def \sTfp {\tilde{\Gamma}^{\mathrm{f}\rightarrow\mathrm{p}}}

\def \sr {\sigr}
\def \si {\sigi}

\def \ii {\mathrm{i}}

\def \mL {\mathcal{L}}

\def \hsig {\hat{\sigma}}

\def \hsigr {\hsig_r}

\def \Jone {\mathrm{J}_1}

\def \bM {\mathbf{M}}
\def \barU {\bar{U}}

\def \pUy {U_y^{\prime}}
\def \pUz {U_z^{\prime}}

\def \mY {\mathcal{Y}}
\def \mZ {\mathcal{Z}}

\def \sigr {\sigma_{r}}
\def \sigi {\sigma_{i}}

\shorttitle{Self-oscillation via an 
electrohydrodynamic instability}
\shortauthor{L. Zhu and H. A. Stone}

\title{Harnessing elasticity to generate self-oscillation via an 
electrohydrodynamic instability}
\author[Lailai Zhu and Howard A. Stone]
{Lailai Zhu$^{1,2,3}$ and Howard A. Stone$^{2}$
  \thanks{Email address for correspondence: hastone@princeton.edu}}

\affiliation{$^1$Department of Mechanical Engineering, National 
University of Singapore, 10 Kent Ridge Crescent, Singapore 119260, 
Singapore\\$^2$Department of Mechanical and Aerospace Engineering, Princeton 
University, Princeton, New Jersey 08544, USA\\$^3$Linn\'{e} Flow Centre and 
Swedish e-Science Research Centre 
(SeRC), KTH 
Mechanics, Stockholm, SE-10044, Sweden}

\begin{document}

\maketitle

\begin{abstract}
Under a steady DC electric field of sufficient strength, 
a weakly conducting 
dielectric sphere in a dielectric solvent with higher conductivity can undergo 
spontaneous spinning (Quincke rotation) through a pitchfork bifurcation. 
We design an object composed of a dielectric sphere and an elastic filament. By 
solving 
an elasto-electro-hydrodynamic (EEH) problem numerically, we uncover an EEH 
instability exhibiting diverse dynamic responses. Varying 
the bending stiffness of the filament, 
the composite object displays three
behaviours: a stationary state, undulatory swimming and steady 
spinning, where the swimming results from a self-oscillatory 
instability through a Hopf bifurcation. By conducting a linear stability 
analysis incorporating an 
elastohydrodynamic model, we theoretically predict the growth 
rates and critical conditions, which agree well with the numerical 
counterparts. 
We also propose a reduced model system consisting of a minimal elastic 
structure which reproduces the EEH instability.
The elasto-viscous response of 
the composite structure is able to transform the pitchfork bifurcation 
into a 
Hopf bifurcation, leading to self-oscillation.
Our results imply a new way of harnessing elastic media to engineer 
self-oscillations, and more generally, to manipulate and diversify the 
bifurcations and the corresponding instabilities. These ideas will be 
useful in designing soft, environmentally adaptive machines.
\end{abstract}

\maketitle

\date{\today}

\section{Introduction}
Active matter has been attracting much 
interest from a broad range of research 
communities~\citep{ramaswamy2010mechanics,cates2011active,
marchetti2013hydrodynamics,needleman2017active}.
At the micron scale, active matter consists of a large number of active 
agents that are able to convert  energy  to achieve 
directed or persistent motions, which include those of living microorganisms, 
synthetic micro-robots, biopolymers such as actin filaments, etc.
The motions of these active agents are attributed to a wide range of 
mechanisms~\citep{lauga2, marchetti2013hydrodynamics, alapan2019microrobotics}, 
e.g. one of the most common strategies adopted by natural and artificial 
micro-swimmers lies in the beating and wiggling of slender structures such as 
cilia and filaments, which are hair-like slender microscale structures  
that play an important role in various biological 
processes~\citep{fawcett1961cilia}, such as 
swimming, pumping, mixing, cytoplasmic streaming, 
etc. The biological 
organelles deliver these functionalities by performing rhythmic,
wave-like motions. 

To achieve persistent motions, cyclic or oscillatory motions are needed, 
yet, the mechanism underlying the emergence of such  
oscillations remains unclear. Two major hypotheses, geometric 
feedback~\citep{brokaw1971bend, brokaw2009thinking, riedel2007synchrony, 
sartori2016dynamic, hines1983three, hilfinger2009nonlinear} 
and ``flutter'' or buckling 
instability~\citep{bayly2016steady,de2017spontaneous, 
ling2018instability,hu2018finite,fatehiboroujeni2018nonlinear}, have been 
raised 
based on theory 
and/or simulations: the first hypothesis assumes that a time-dependent dynein 
activity (switching on/off or modulation) is necessary to trigger 
the oscillations; the second one suggests that a steady
point force or force distributions acting along the axial direction 
of a flexible filament can trigger 
its oscillatory motion through a ``flutter'' or buckling 
instability.
These forces are in fact called the ``follower force'' in the mechanics 
literature~\citep{pfluger1950stabilitatsprobleme, 
ziegler1952stabilitatskriterien, herrmann1964stability}. Because
the follower force was initially invented theoretically and 
assumed to be always tangential to the slender structure regardless of its 
time-dependent deformation, it was demonstrated only mathematically and has 
been considered impractical~\citep{koiter1996unrealistic}. However, it was 
recently
realised experimentally on a metre-scale 
rod~\citep{bigoni2018flutter}.

To drive the oscillations of artificial cilia and filaments of micron scale, 
different methods that exploit 
magnetic~\citep{dreyfus2005microscopic,singh2005synthesis,
evans2007magnetically,
livanovivcs2012magnetic, hanasoge2017asymmetric,huang2019adaptive}
, electrostatic~\citep{den2008artificial}, piezoelectric~\citep{oh2009bio},  
 optical~\citep{van2009printed} and hydrogel-based 
actuations~\citep{sidorenko2007reversible,masuda2013self} have been developed. 
Nonetheless, these practises relied on a time-dependent power source, except 
for 
the self-oscillation of polymer brushes triggered by
the Belousov-Zhabotinsky reaction~\citep{masuda2013self}. This
reaction-based beating shares with other biological processes, 
such as mammalian  otoacoustic 
emissions~\citep{gold1948hearing,kemp1979evidence} and 
glycolysis~\citep{sel1968self} the same 
feature: self-oscillation, that is generating and sustaining a periodic 
motion 
based on a power source without a corresponding 
periodicity~\citep{jenkins2013self}.

In our recent work~\citep{zhu2019propulsion}, we proposed a 
chemical-reaction-free and follower-force-free strategy to engineer the 
self-oscillations of artificial structures by employing a time-independent, 
uniform electric field. We reported an 
elasto-electro-hydrodynamic (EEH) instability based on the Quincke rotation 
(QR) instability, and utilised it to drive various motions of an object 
composed of a dielectric spherical particle with an attached 
elastic filament. In this work, we will present in detail the setup and the 
mathematical description of the new EEH problem. First, we numerically
solve the system  coupling the  electrohydrodynamics of the particle in a 
dielectric fluid and the elastohydrodynamics of the filament in a viscous 
fluid. We 
identify the emergence of the EEH instability  that produces the
self-oscillation of the composite object. The oscillations in turn
cause the object to translate. Then, we perform a linear stability 
analysis (LSA) incorporating an elastohydrodynamic model to predict  
the onset of self-oscillatory instability. Finally, we propose a minimal model
that can reproduce the similar EEH instability.

We describe the setup and governing equations
of the EEH problem in \S~\ref{sec:setup_math}, and demonstrate the numerical 
results in \S~\ref{sec:results}. The elastohydrodynamic model and LSA are 
shown in \S~\ref{sec:linear}, followed by \S~\ref{sec:minimal} illustrating 
the 
minimal model. Finally,
we conclude our observations and provide some discussions in 
\S~\ref{sec:conclusions}.

\section{Problem setup and mathematical formulations}\label{sec:setup_math}
\begin{figure}
\begin{center}
\includegraphics[scale=0.8]{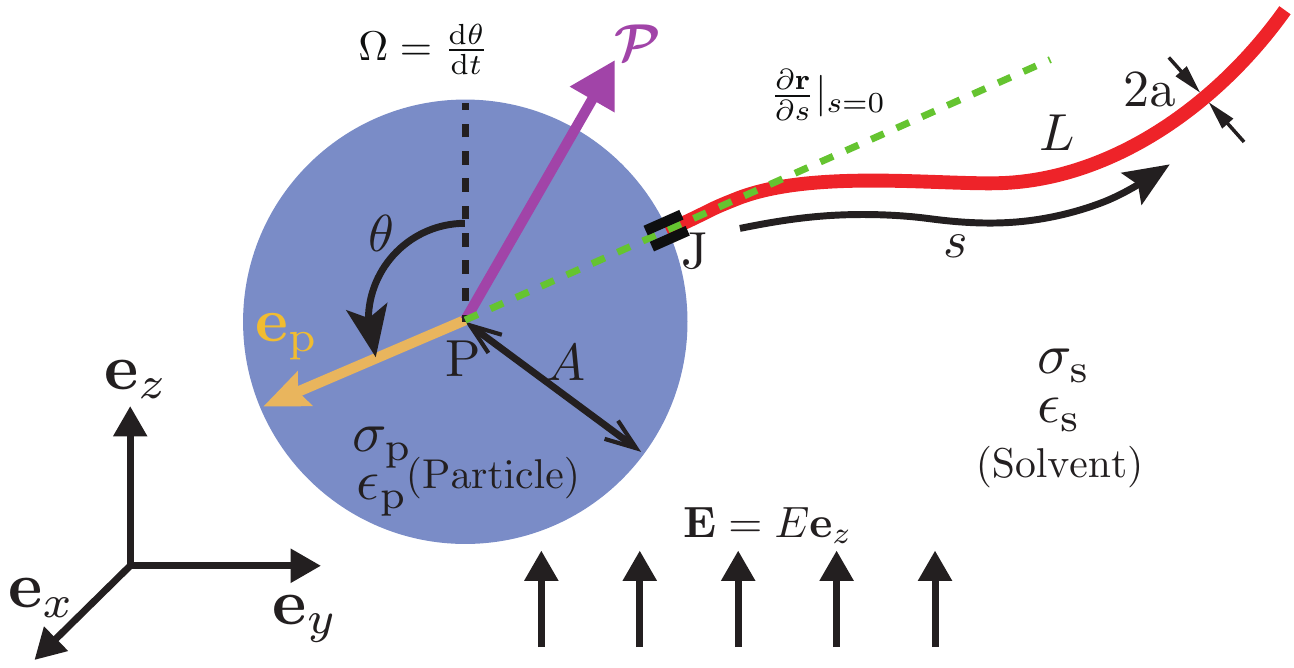}
\end{center}
\caption{Schematic of the setup: a dielectric spherical particle of radius $A$ 
attached with a flexible filament of contour length $L$ is
exposed to a steady uniform electrical field $\bE = E\be_z$. The composite 
object's motion, orientation $\ortp$ 
and induced dipole $\bP$ are all constrained to the $yz$-plane, and $\ortp$ is 
described by the angle $\theta$ with respect to $\be_z$.}
\label{fig:sketch}
\end{figure}

We consider a weakly conducting dielectric spherical particle of radius $A$, 
which has attached an inextensible elastic filament of contour length $L$. The 
filament is 
cylindrical with a constant cross-section of radius $a$, and its slenderness  
is 
$\esl = a/L \ll 1$. We fix $\esl = 0.01$ in 
this work. The 
composite object is subject to a 
time-independent uniform electric field $\bE = E \be_z$ (see 
figure~\ref{fig:sketch}), where $\mathbf{e}_z$ is the $z$-direction basis vector 
of the laboratory coordinates system $\mathbf{e}_{xyz}$. The centreline of the 
filament is described 
by $\br \lp s, t\rp$, where $s$ indicates the arclength.   The  
base J ($s=0$) of the filament is clamped at the particle surface, namely, 
the tangent vector $\partial \br/\partial s |_{s=0}=-\ortp$ at the base always 
passes through the particle centre P, regardless of its deformation and 
the orientation vector $\ortp$ of the particle.
The size ratio between the particle and filament is $\alpha = A/L$.
We consider only the bending deformation of the filament with a 
bending stiffness of $D = \pi a^4 Y/4$, where $Y$ denotes Young's modulus.

The composite object is immersed in a dielectric solvent 
fluid with dynamic viscosity $\mu$. The electrical conductivity and absolute 
permittivity of the solvent are
$\ss$ and $\es$, respectively, and those of the particle are $\sp$ and $\ep$;
$R=\sp/\ss$ and $S=\ep/\es$ indicate the ratios. The terms $\taus = \es/\ss$ and
$\taupart = \ep/\sp$ denote the charge relaxation time of the solvent and 
particle, respectively. These electrical properties are important to the 
induced QR electrohydrodynamic instability that is critical to the dynamics in 
this paper. Their values are based on
experiments~\citep{brosseau2017electrohydrodynamic}, where $R=2.3 \times 
10^{-7}$ and $S=0.84$ are fixed in this work.
Though the filament will also be polarised like the particle, the induced 
electric torque on the filament will be much weaker than that on the particle 
(see \S~\ref{sec:conclusions} for a detailed discussion). We thus do not 
consider the electrohydrodynamics of the filament in this work.

\subsection{Assumptions}
The numerical simulations are carried out by invoking several assumptions.
Motivated by biomimetic applications at the micron 
scale, we neglect the inertia of the fluid and particle. The fluid motion 
is therefore governed by the Stokes equations, and the particle satisfies 
instantaneous force- and torque-free conditions. The movement of the composite 
object is constrained to be planar, such that the particle centre P and 
filament 
position $\br(s, t)$ are in the $yz$-plane. 

We adopt the local resistive-force theory~\citep{batchelor1970slender} to 
calculate the hydrodynamic forces on the filament. We further ignore the
hydrodynamic interactions between the particle and the filament. 
In the elastohydrodynamic model developed for LSA, we also assume that the 
filament undergoes weak deformation.

\subsection{Electrohydrodynamics of the particle}
\begin{figure}
\begin{center}
\includegraphics[width=0.6\textwidth]
{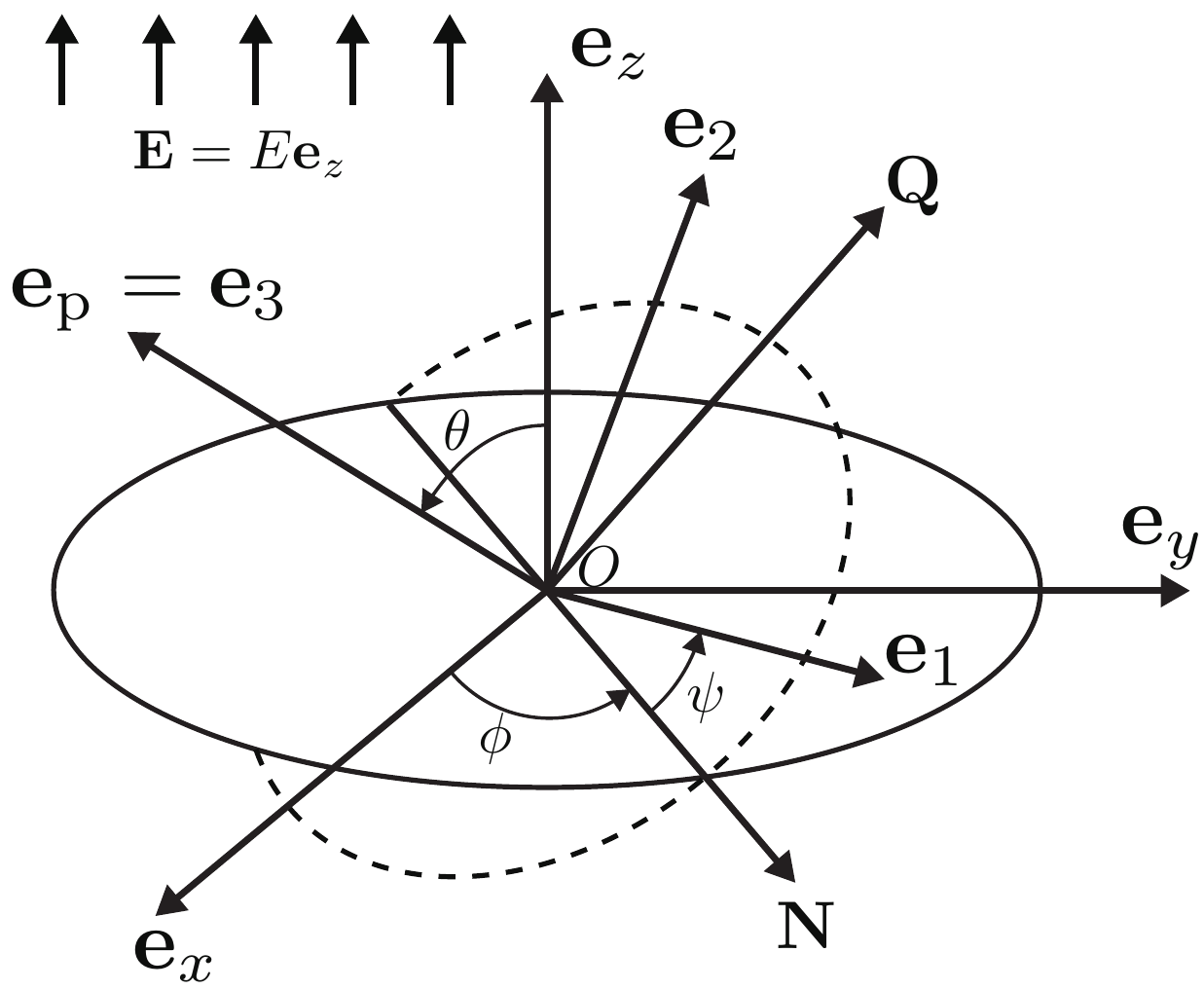}
\end{center}
\caption{The reference frame $\be_{123}$ that rotates and translates with the
particle,
the orientation $\ortp$ of the composite object
coincides with $\be_3$. The proper Euler angles $[\theta, \phi, 
\psi]$ are adopted to describe the orientation $\ortp$, where $\bN$ denotes the 
nodal line direction and $\bQ = \be_3 \times \bN$. We note that a graphical 
error occurred in a similar figure in our related 
work~\citep{zhu2019propulsion}, where $\psi$ ranges from $\mathbf{N}$ but 
erroneously to $\mathbf{e}_y$.
}
\label{fig:euler-sketch}
\end{figure}

When a dielectric particle in a dielectric 
solvent is exposed to an electric field, the interface of the particle will be 
polarised. The total
induced dipole $\bPtt$ consists of an instantaneous part $\bPinf$
and a retarding part $\bP$, \textit{viz.} $\bPtt = \bPinf + \bP$. Both
vectors are defined by three components, $\mPinf_i$ and $\mP_i$ ($i=1...3$) in 
the reference frame $\be_{123}$ that rotates with the particle 
(see 
figure~\ref{fig:euler-sketch}). 
For a homogeneous spherical particle, its 
Maxwell-Wagner polarisation time $\taumw$, and low- and high-frequency
susceptibilities, $\chi^0$ and $\chi^{\infty}$, respectively, are isotropic, 
hence the $i$-th 
component of the instantaneous dipole $\bPinf$ is 
\begin{align}\label{eq:mPinf}
 \mPinf_i = \chi^{\infty} E_i.
\end{align}
In the rotating reference frame of the particle, the retarding  dipole 
$\bP$ is 
governed 
by~\citep{tsebers1980internal, cebers2000electrohydrodynamic} 
\begin{align}
 \frac{\partial  \mP_i}{\partial  t} = -\frac{1}{\taumw}\left[ \mP_i - \lp 
\chi^{0} - 
\chi^{\infty} 
\rp E_i\right],
\end{align}
where  
\begin{align}\label{eq:kappa}
 \kappa=\frac{R+2}{S+2}
\end{align}
 and $\taumw = \taus / \kappa$. It is 
well known that 
when the charge 
relaxation time $\taupart$ of the particle is larger than that  of the 
solvent $\taus$, \textit{i.e.}, $R/S<1$, $\bP$ is oriented opposite to
 the electric field. This 
directional misalignment is the necessary condition for the electro-rotation of
the particle, the so-called Quincke rotation~\citep{quincke1896ueber}, 
which 
occurs when, in addition, the strength $E$ of the electric field is above
a critical value $\Ecri$ derived theoretically 
as~\citep{jones1984quincke,brosseau2017electrohydrodynamic} (see 
appendix~\ref{sec:append})
\begin{align}\label{eq:Ecri}
 \Ecri = \sqrt{\frac{2 \ss \mu (R+2)^2}{3 \es^2(S-R)}}.
\end{align}

We do not consider the hydrodynamic interactions between
the spherical particle and filament, hence the dynamics of the particle can be 
obtained by using its translational and rotational mobility factors. 
Assuming that the particle rotates at angular velocity $\Omp$ about its centre 
P,
which translates at velocity $\Up$, the force and torque balances on 
the particle give
\begin{subequations}
\begin{align}
 \bFfp - \beta_{\mathrm{drag}} \Up  & = \mathbf{0}, \\
 \bTfp +  \bTel - \gamma_{\mathrm{drag}} \Omp & = \mathbf{0}, 
\label{eq:torque-free}
\end{align} 
\end{subequations}
where $\bFfp$ denotes the force exerted by the filament on the particle, 
$\bTfp$ the torque with respect to the particle centre P, and 
$\beta_{\mathrm{drag}} 
= 
6\pi \mu A$
and $\gamma_{\mathrm{drag}} = 8\pi \mu A^3$ are the translational and 
rotational
drag 
coefficients
of a sphere in the creeping flow, respectively. Also,       
 $\bTel$ is the electric 
torque on the particle with respect to its centre P, that is
\begin{align}
 \bTel = \bPtt \times \bE =  \bPinf \times \bE + \bP \times \bE = \bP \times 
\bE,
\end{align}
where $\bPinf \times \bE \equiv \mathbf{0}$ for an isotropic sphere
because 
$\mPinf_i$ linearly scales with $E_i$ in each direction by the same factor 
$\chi^{\infty}$ (see equation~(\ref{eq:mPinf})). It is worth noting that
$\bPinf \times \bE \neq \mathbf{0}$ for ellipsoidal particles, where the 
factor $\chi^{\infty}$ is direction dependent
\citep{cebers2000electrohydrodynamic,brosseau2017electrohydrodynamic}. 
The translation of the particle is driven by the 
elastic force exerted by the filament, which is balanced by the viscous drag, 
while the rotational motion of the particle is determined by the balance 
between the elastic, electric and hydrodynamic torques.

The orientation of the particle $\ortp$ is defined as the direction 
from the filament base J towards the particle centre P, where $\be_3$
of the particle-based reference frame coincides with $\ortp$. We have found it 
convenient to 
use the
proper Euler angles $[\theta, \phi, \psi]$, see 
figure~\ref{fig:euler-sketch}. Here,
$\bP$ is decomposed into $\bP = \mP_3 \be_3 + \mP_N \bN + 
\mP_Q \bQ$,
where $\bN$ indicates the nodal line direction and $\bQ = \be_3 \times \bN$.
This decomposition applies to other vectorial variables such as $\bE$.
We constrain  $\bP$ onto the $yz$-plane, hence $\phi=\psi\equiv 0$ and 
$\theta$ is the only angle indicating the orientation $\ortp$; 
additionally, $\mP_N = 0$ and $\be_x = \bN$. For the sake of 
completeness, we first derive the governing equations for a general 
situation without these constraints.
 
Using the torque-free condition equation~(\ref{eq:torque-free}), we obtain the 
governing equations for $[\theta, \phi, \psi]$,
\begin{subequations}\label{eq:euler-dim}
\begin{align}
 \frac{\partial \theta}{\partial t} & = \frac{1}{\gamma_{\mathrm{d}}} 
\lp\Tfp_N + E_3 \mP_Q 
- E_Q \mP_3 \rp, \\
\frac{\partial \phi}{\partial t} & = \frac{1}{\gamma_{\mathrm{drag}}} 
\sin{\theta} 
\lp -E_3 
\mP_N 
+ \Tfp_Q \rp, \\
\frac{\partial \psi}{\partial t} & = \frac{1}{\gamma_{\mathrm{drag}}} 
\sin{\theta}\lp E \mP_N 
- \Tfp_Q \cos{\theta}\rp,
\end{align} 
\end{subequations}
where $E_3 = E\cos{\theta}$ and $E_Q = E\sin{\theta}$.
The governing equations for $[\mP_N, \mP_Q, \mP_3]$ 
are~\citep{cebers2000electrohydrodynamic}
\begin{subequations}\label{eq:dipole-dim}
\begin{align}
\frac{\partial \mP_N}{\partial t} + \frac{\partial  \psi}{\partial t} \mP_Q & 
= -\frac{1}{\taumw} \mP_N, \label{eq:PN} \\
\frac{\partial \mP_Q}{\partial t} - \frac{\partial \psi}{\partial t} \mP_N & 
= 
-\frac{1}{\taumw} \left[ \mP_Q - \lp \chi^0 - \chi^{\infty} \rp E_Q \right],  
\label{eq:PQ}\\
\frac{\partial \mP_3}{\partial t} & = - \frac{1}{\taumw} \left[ \mP_3 - \lp 
\chi^0 
- \chi^{\infty} \rp E_3 \right]. \label{eq:P3}
\end{align} 
\end{subequations}

We choose the charge relaxation time of the solvent $\taus$ as the 
characteristic time, $L/\taus$ the characteristic velocity, and $\Ecri$ and 
$D/(L 
\Ecri)$ the characteristic strength of the electrical field and polarisation 
dipole, respectively. Using $\;\bar{}\;$ to indicate the  dimensionless 
variables 
hereafter, the dimensionless equations for the Euler angles
$[\theta, \phi, \psi]$ are
\begin{subequations}\label{eq:non-euler}
\begin{align}
 \frac{\partial \theta}{\partial \nont} & = \frac{1}{\bareta} \lp\nTfp_N + 
\nE_3 \nP_Q 
- \nE_Q \nP_3 \rp,  \label{eq:non-theta} \\
\frac{\partial \phi}{\partial \nont} & = \frac{1}{\bareta \sin{\theta}} \lp 
-\nE_3 \nP_N 
+ \nTfp_Q \rp,  \label{eq:non-phi} \\
\frac{\partial \psi}{\partial \nont} & = \frac{1}{\bareta \sin{\theta}}\lp 
\nE \nP_N  
- \nTfp_Q \cos{\theta}\rp, \label{eq:non-alpha}
\end{align} 
\end{subequations}
as derived in \citet{cebers2000electrohydrodynamic}
in the absence of the elastic torque 
$\bar{\boldsymbol{\Gamma}}^{\mathrm{f}\rightarrow\mathrm{p}}$. Here,
\begin{align}\label{eq:bareta}
 \bareta = \alpha^3 \barmu,
\end{align}
with
\begin{align}\label{eq:barmu}
\barmu = \frac{8\pi \mu L^4}{D \taus} 
\end{align}
defined as the elasto-electro-viscous (EEV) parameter. 
The dimensionless governing equations 
for $[\nP_N, \nP_Q, \nP_3]$ following from equation~(\ref{eq:dipole-dim}) are
\begin{subequations}
 \begin{align}
  \frac{\partial \nP_N }{\partial \nont} & = -\kappa \nP_N -\frac{\partial 
\psi}{\partial \nont}\nP_Q, \label{eq:non-PN} \\
 \frac{\partial \nP_Q}{\partial \nont} & = -\kappa \lp\nP_Q + \kappa \bareta 
\nE_Q \rp + \frac{\partial \psi}{\partial t} \nP_N,  \label{eq:non-PQ} \\
\frac{\partial \nP_3}{\partial \nont} & = -\kappa \lp \nP_3 + \kappa \bareta 
\nE_3 \rp, \label{eq:non-P3}
 \end{align}
\end{subequations}
where $\kappa = (R+2)/(S+2)$ as defined in equation~(\ref{eq:kappa}). 
We slightly perturb the instantaneous polarisation $\bar{\bP}^{\infty}$ 
and use it as the initial value  
$\bar{\bP}_{\mathrm{ini}}$ of $\bar{\bP}$, where
\begin{subequations}\label{eq:bp_perturb}
 \begin{align}
  \bar{\bP}^{\infty} & = \frac{\bareta \kappa^2 \barE}{\kappa-(R-1)/(S-1)}\lp 
\be_Q \sin\theta  + \be_3 \cos\theta \rp, \\
  \bar{\bP}_{\mathrm{ini}} & = \bar{\bP}^{\infty}  + 
\epsilon_{\mP}|\bar{\bP}^{\infty}|,
 \end{align}
\end{subequations}
with $\epsilon_{\mP}=\mathcal{O}(10^{-4}) - 
\mathcal{O}(10^{-3})$. 

The dimensionless  force- and torque-free conditions 
are
\begin{subequations}\label{eq:force_torque_non}
\begin{align}
 \nonbFfp - 3\alpha \barmu \nonbUp /4 & = \mathbf{0}, \label{eq:force-free-non} 
\\
 \nonbTfp + \nonbP \times \nonbE  - \bareta \nonbOmega & = 
\mathbf{0}.\label{eq:torque-free-non} 
\end{align} 
\end{subequations}

Since we constrain the motion of the composite object and the induced dipole 
$\bar{\bP}$ to the $yz$-plane, we solve
equations~(\ref{eq:non-theta}), (\ref{eq:non-PQ}) and (\ref{eq:non-P3}) for 
$\theta$, 
$\nP_Q$ and $\nP_3$, where the last term $\frac{\partial \psi}{\partial t} 
\nP_N$ in equation~(\ref{eq:non-PQ}) vanishes.

\subsection{Elastohydrodynamics of the filament}
We describe here the elastohydrodynamic equations for the 
filament. 
By employing the slender body theory (SBT) considering the leading-order local 
drag~\citep{batchelor1970slender}, 
the relation between the velocity $\br_t$ of the filament centreline and 
the 
force per unit length exerted
by the fluid onto the filament $\bf(s,t)$ 
is
\begin{align}\label{eq:local}
 8 \pi \mu \lp \br_t - \bUinf \rp  = c \lp \bI + \br_s  \br_s \rp \cdot 
\bf,
\end{align}
where $\bUinf$ is the underlying flow velocity (background or imposed 
flow velocity) at $\br(s,t)$ and $\bUinf = 
\mathbf{0}$ in this work; the 
subscripts $t$ and $s$ denote the partial derivatives  with respect to $t$ and 
$s$, respectively and
\begin{align}\label{eq:c}
c = 1+2 \log{\esl}<0.
\end{align}
The filament is assumed to be described by the Euler–Bernoulli 
constitutive law, and because the elastic force balances the hydrodynamic force 
anywhere on the centreline, we obtain
\begin{align} \label{eq:bf}
 \bf(s) = -\lp T(s) \br_s \rp_s + D \br_{ssss},
\end{align}
where $T(s,t)$ denotes the line tension, which acts as a Lagrangian multiplier 
to 
guarantee the inextensibility of the filament, \textit{i.e.}, 
$\br_s \cdot \br_s \equiv 1$.

By substituting 
equation~(\ref{eq:bf}) into equation~(\ref{eq:local}), and choosing $L$ and 
$D/L^2$ 
as the 
characteristic length and force, respectively, we obtain the dimensionless 
equations for $\nonbr(\nons,\nont)$,
\begin{align}\label{eq:fila_pos}
 \barmu \nonbr_{\nont} = -2c \nonT_{\nons} \nonbr_{\nons} - c \nonT 
\nonbr_{\nons\nons} + c \nonbr_{\nons\nons\nons\nons} + c 
\lp \nonbr_{\nons}\cdot \nonbr_{\nons\nons\nons\nons} 
\rp\nonbr_{\nons}.
\end{align}
The 
dimensionless  equation
for $\nonT \lp \nons \rp$ reads,
\begin{align}\label{eq:fila_T}
 2c \nonT_{\nons\nons} - c \nonT\nonbr_{\nons\nons}\cdot 
\nonbr_{\nons\nons} = -7c\nonbr_{\nons\nons}\cdot 
\nonbr_{\nons\nons\nons\nons} - 6c 
\nonbr_{\nons\nons\nons} \cdot \nonbr_{\nons\nons\nons} - \barmu \bein 
\lp 1 - \nonbr_{\nons}\cdot \nonbr_{\nons} \rp,
\end{align}
where the last term on the right-hand side $- \barmu \bein 
\lp 1 - \nonbr_{\nons}\cdot \nonbr_{\nons} \rp$ is an extra (numerical) 
penalisation term 
introduced~\citep{tornberg2004simulating,li2013sedimentation} to 
preserve the local inextensibility constraint $\nonbr_{\nons} \cdot 
\nonbr_{\nons} \equiv 1$; $\bein=100$ is adopted in our simulations.
The boundary conditions (BCs) for $\nonbr(\nons,\nont)$ and 
$\nonT(\nons,\nont)$ 
at the free end 
$\nons=1$ are
\begin{subequations} \label{eq:BC_s1}
 \begin{align}
  \nonbr_{\nons\nons} & = \nonbr_{\nons\nons\nons} = 
\mathbf{0}, \label{eq:BC_s1_r}\\
  \nonT & = 0.
 \end{align}
\end{subequations}
The BCs at the clamped end $\nons=0$  couple the elastohydrodynamics and
electrohydrodynamics, as will be described next.

\subsection{Elasto-electro-hydrodynamic coupling}\label{sec:eeh_coup}
The electrohydrodynamics of the dielectric particle in a dielectric solvent
and the elastohydrodynamics of the flexible filament in a viscous fluid are 
coupled via, first the BCs of $\nonbr(\nons,\nont)$ and 
$\nonT(\nons,\nont)$ at the filament base $\nons=0$, and second the 
elastic force $\nonbFfp(\nont)$ and torque $\nonbTfp(\nont)$ exerted by the 
filament 
on the particle (equation (\ref{eq:force_torque_non}).

The BCs at the filament base $\nons=0$  are
\begin{subequations}\label{eq:BC_s0}
 \begin{align}
  \nonbr & = \nonbxp + \alpha \nonbr_{\nons},  \label{eq:nonbr_s0} \\
  \nonbr_{\nons} & = -\ortp, \label{eq:nonbrs_s0}
 \end{align}
\end{subequations}
where $\nonbxp (\nont)$ denotes the dimensionless position of the particle 
centre P. 
Equations~(\ref{eq:nonbr_s0}) and ~(\ref{eq:nonbrs_s0}) imply, respectively, 
that the filament base $\nons=0$ is exactly
on the particle surface, and the filament tangent vector at $\nons=0$ always 
passes
through the particle centre. Moreover,
$\nonbxp(\nont)$ and $\ortp(\nont)$ are connected to  the particle kinematics 
through
\begin{subequations}\label{eq:par-kin}
 \begin{align}
 \frac{\d \nonbxp}{\d \nont} & = \nonbUp, \\
 \frac{\d \ortp}{\d \nont} & = \bar{\boldsymbol{\Omega}} \times \ortp,
 \end{align}
\end{subequations}
where $\nonbUp(\nont)$ is linked to equation~(\ref{eq:force-free-non}) and 
$\bar{\boldsymbol{\Omega}}(\nont)$ to
equation~(\ref{eq:non-euler}).
The coupling is completed by the computation of $\nonbFfp$ and $\nonbTfp$,
\begin{subequations}\label{eq:force-torque-non}
\begin{align}
\nonbFfp & = \left[-\nonbr_{\nons\nons\nons} + 
\nonT\nonbr_{\nons}\right]|_{\nons=0}, \label{eq:non_elastic_force}\\
\nonbTfp & = \left[\nonbr_{\nons} \times \lp \nonbr_{\nons\nons} - \alpha 
\nonbr_{\nons\nons\nons} \rp\right]|_{\nons=0}.\label{eq:non_elastic_torque}
\end{align} 
\end{subequations} 

For completeness, we write the BC for the tension $\nonT$ at the filament 
base 
$\nons=0$ 
\begin{align}
 2 c \nonT_{\nons} + 6c \nonbr_{\nons\nons} \cdot \nonbr_{\nons\nons\nons} = 
- \barmu \nonbr_{\nons} \cdot \nonbr_{\nont}.
\end{align}

\section{Numerical results}\label{sec:results}
In the original QR phenomenon (without a filament),  the particle rotates 
when the dimensionless electric field is above $1$, namely, $\barE \geq 1$.
We hereby investigate the influence of the bending stiffness
of the filament by varying $\barmu$, where we fix the electric 
field $\barE = 1.5$ at which an individual particle 
undergoes steady QR. 
We fix the 
size ratio $\alpha=0.3$ in this section.

\begin{figure}
\begin{center}
\hspace{0cm}\includegraphics[width=1\textwidth]
{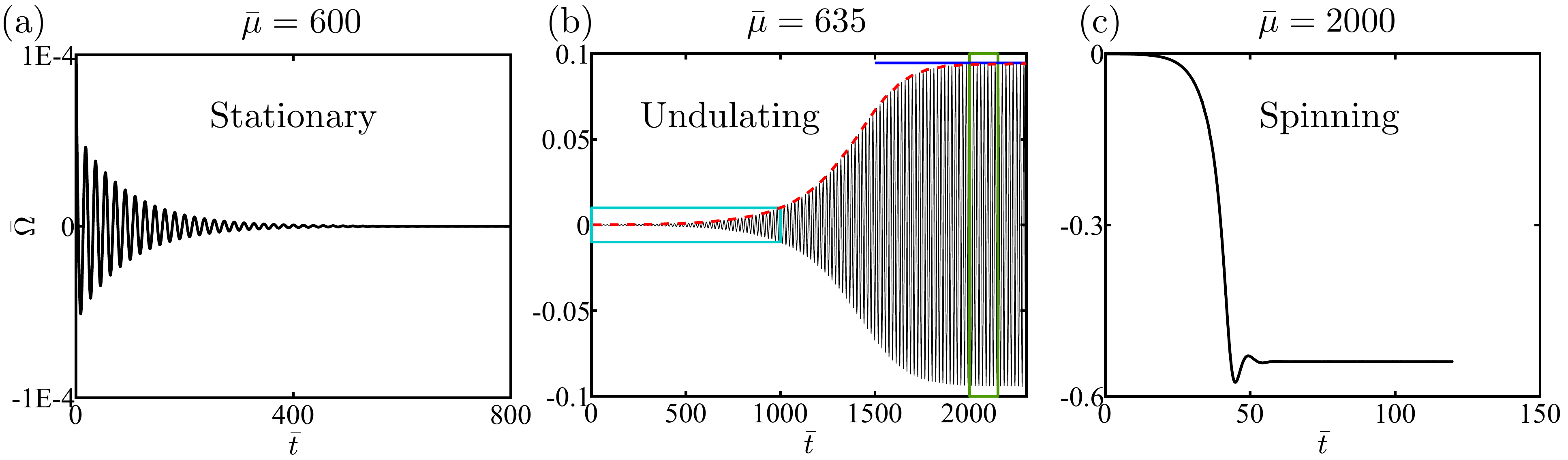}
\end{center}
\caption{$\barmu$-dependent time evolution of the 
rotational velocity $\nonOmega \lp \nont \rp$ for (a) $\barmu=600$, (b) 
$\barmu=635$ and (c) $\barmu=2000$  when $\barE = 
1.5$. Their 
corresponding equilibrium configurations are stationary, 
undulating and steady spinning, respectively. Note that 
$\nonOmega_{\nont=0}$ is not necessarily zero because the induced dipole 
$\bar{\bP}$ is slightly perturbed at $\nont=0$, see 
equation~(\ref{eq:bp_perturb}); moreover, (a) and (b) have strikingly  
different scales for $\nonOmega$.  }
\label{fig:omg_v_time_1}
\end{figure}

\begin{figure}
\begin{center}
\hspace{0cm}\includegraphics[width=1\textwidth]
{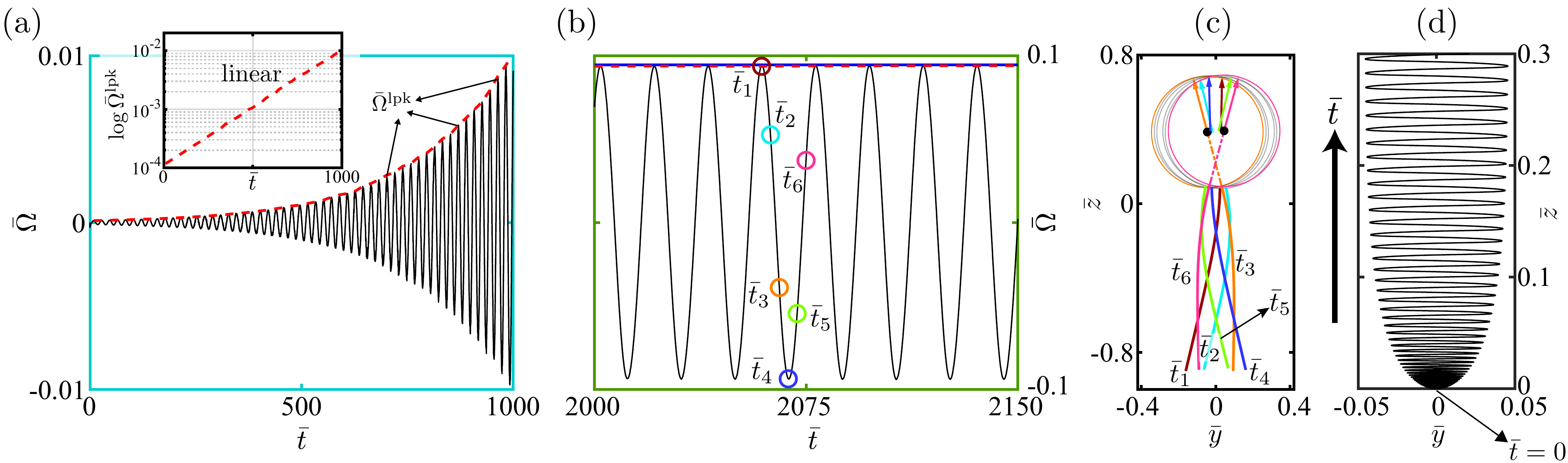}
\end{center}
\caption{
(a) Highlighted cyan domain of 
figure~\ref{fig:omg_v_time_1}b indicating the initial rapidly growing period of 
$\nonOmega(\nt)$, for $\barmu=635$ and $\barE = 1.5$. The red 
curve denotes the local peak $\nonOmegalpk$, and the inset of (a) shows the 
linear dependence of $\log{\nonOmegalpk}$ on $\nont$. (b) Highlighted 
green domain of figure~\ref{fig:omg_v_time_1}b corresponding to the 
time-periodic response of $\nonOmega(\nt)$, where consecutive time instants 
$\nont_i$ ($i=1,...,6$) within a period are marked. (c) Particle-filament 
configurations  at $\nont_i$. (d) Trajectory of 
the particle centre within $\nont 
\in [0, 1940]$.}
\label{fig:zoom_in_mu635}
\end{figure}

We observe that the composite object exhibits three $\barmu$-dependent 
scenarios, indicated by the time evolution of the rotational velocity 
$\nonOmega$ shown in 
figure~\ref{fig:omg_v_time_1}. When $\barmu=600$ 
(figure~\ref{fig:omg_v_time_1}a), $\nonOmega$ decays dramatically and 
eventually becomes zero, indicating that the object relaxes to a stationary 
state.
Increasing $\barmu$ to $635$ (figure~\ref{fig:omg_v_time_1}b), the 
time evolution of $\nonOmega$ 
features two phases: in the initial phase (cyan domain), it grows 
rapidly due to self-oscillation; in the second  phase (green domain), it 
reaches a 
time-periodic state with a 
constant amplitude of approximately $0.1$.
The third type of response is illustrated by $\barmu=2000$, where $\nonOmega$ 
eventually approaches a steady value around 
$-0.6$.

We further scrutinise the $\barmu=635$ case. The close-up views of the 
initially rapidly growing
phase (cyan domain) and saturated time-periodic phase (green domain) are shown 
in figure~\ref{fig:zoom_in_mu635}a and b, respectively. The red curve 
connecting the local peaks $\nonOmegalpk$ of $\nonOmega$ implies an
exponential growth of $\nonOmega$ in time. This trend is confirmed by the
linear relationship between $\log{\nonOmegalpk}$ and $\nont$ shown in the
inset of figure~\ref{fig:zoom_in_mu635}a. 
The time-periodic phase enlarged in figure~\ref{fig:zoom_in_mu635}b 
reveals its sinusoidal-like variation characterised  by fore-aft temporal 
symmetry. Six times within one 
period of this phase are marked, with their corresponding positions and 
orientations of the particle, and the profiles of the filament
depicted in figure~\ref{fig:zoom_in_mu635}c. The oscillating  particle 
drives the filament to wiggle, because the filament
is clamped onto the particle. The wiggling filament provides thrust
to the whole object, as a natural resemblance to a biological appendage.
Consequently, the object achieves locomotion, following a wave-like
trajectory (figure~\ref{fig:zoom_in_mu635}d). The 
wavy path is tightly packed near $\nont=0$, implying the slow motion of the 
object undergoing small-amplitude oscillation in the initial phase.

\begin{figure}
\begin{center}
\hspace{0em}\includegraphics[width=1\textwidth]
{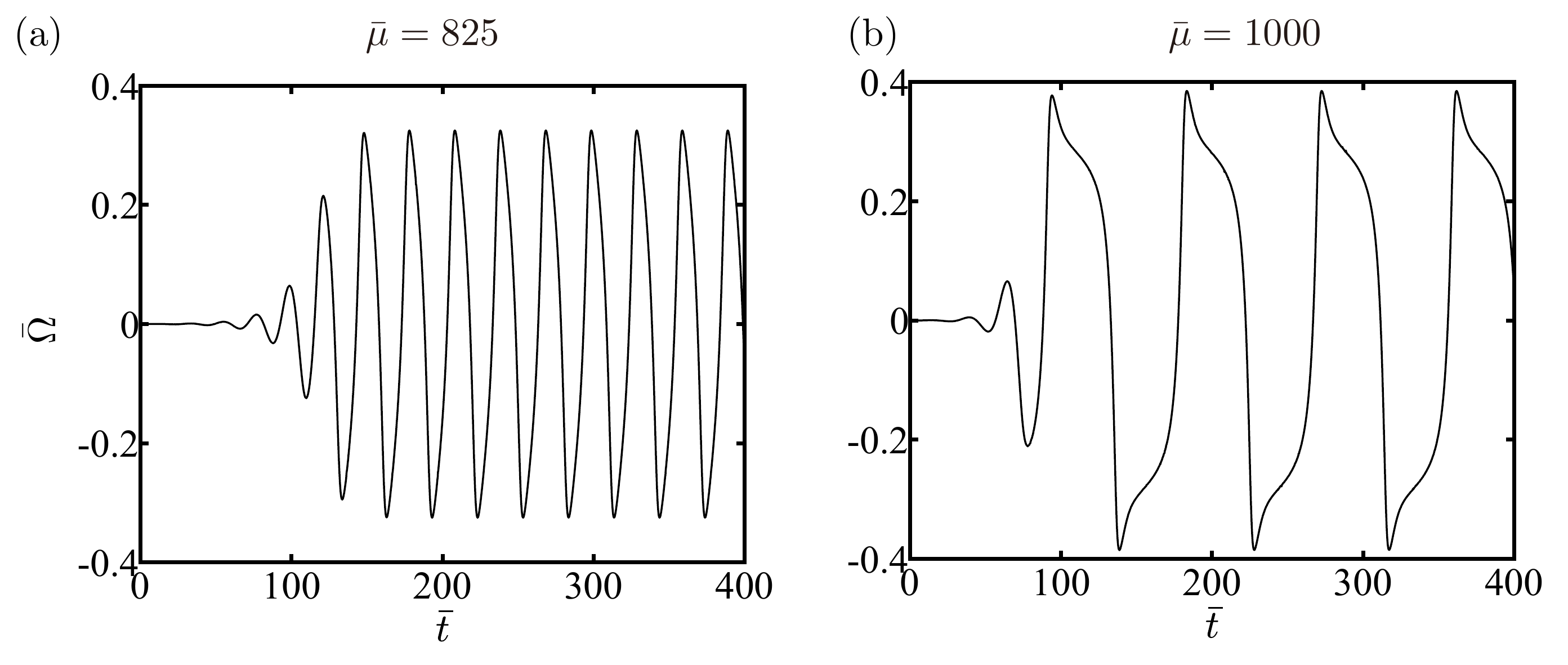}
\end{center}
\caption{Time evolution of the
rotational velocity $\nonOmega \lp \nont \rp$ when  $\barE = 
1.5$ for (a) $\barmu=825$ and (b) $\barmu=1000$.}
\label{fig:omg_v_time_2}
\end{figure} 
 
We observe that when $\barmu$ lies in the self-oscillating regime,
the time evolution of $\nonOmega$ varies with $\barmu$. As shown in 
figure~\ref{fig:omg_v_time_2} for $\barmu=825$
and $1000$, for a larger $\barmu$ it takes fewer time 
periods for the perturbation to reach its time-periodic state.
In addition, that state clearly breaks fore-aft 
symmetry with increasing $\barmu$.

\begin{figure}
\begin{center}
\hspace{0em}\includegraphics[width=1\textwidth]
{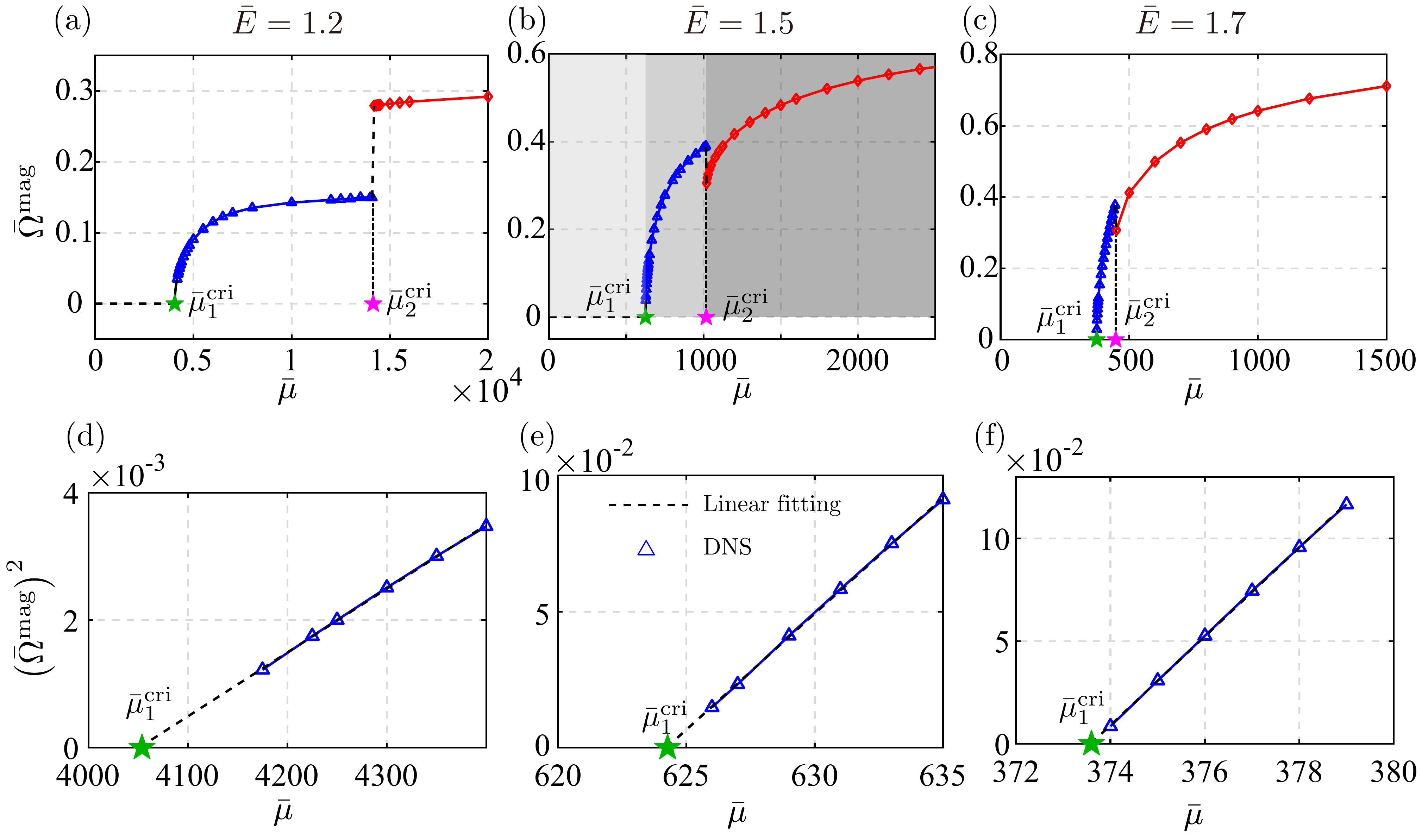}
\end{center}
\caption{Amplitude $\nonOmegaamp$ of the  
rotational velocity as a function $\barmu$ for  (a) $\barE=1.2$, (b) $1.5$ and 
(c) $1.7$. The three $\barmu$-dependent 
regimes, stationary (dashed lines), undulating (triangles) and steady spinning 
(diamonds) of the composite object are separated by two thresholds
$\barmucrione$ and $\barmucritwo$. (d), (e) and (f)  
show the linear variation of $\lp \nonOmegaamp \rp^2$ in $\barmu$ in close 
proximity to $\barmucrione$ for $\barE=1.2$, $1.5$ and 
$1.7$, respectively.}
\label{fig:omg_mu_alpha03_varyE}
\end{figure} 
 
We next investigate the critical $\barmu$ values that separate the three
regimes corresponding to the stationary, undulating and steady spinning
states. Figure~\ref{fig:omg_mu_alpha03_varyE} displays the rotational 
velocity magnitude $\nonOmegaamp$ versus $\barmu$ for $\barE=1.2$ (a), $1.5$ 
(b) 
and $1.7$ (c). When 
$\barmu \leq \barmucrione$, $\nonOmegaamp=0$ 
represents the fixed-point solution;
when $\barmu \geq \barmucrione$, the non-zero 
$\nonOmegaamp$ representing 
the constant spinning speed corresponds to the asymmetric 
fixed-point solution; when $\barmu \in \lp \barmucrione, \barmucritwo\rp$,
$\nonOmegaamp$ indicates the magnitude of the oscillating 
$\nonOmega$ when it reaches a time-periodic state. We plot $\lp 
\nonOmegaamp \rp^2$ as a function of $\barmu$ in close proximity to 
$\barmucrione$ in figure~\ref{fig:omg_mu_alpha03_varyE}d-f. The 
linear dependence of $\lp 
\nonOmegaamp \rp^2$ on $\barmu$ implies that the 
instability occurs at $\barmucrione$ through a Hopf
bifurcation from where  a limit-cycle 
solution emerges. Moreover, the $\nonOmegaamp(\barmu)$ profile also 
indicates the supercritical nature of the Hopf bifurcation. On the other hand, 
a sudden jump of 
$\nonOmegaamp$ at $\barmucritwo$ signifies a secondary 
bifurcation where the limit cycle
shrinks to a fixed point or vice versa.

\begin{figure}
\begin{center}
\hspace{0em}\includegraphics[width=1\textwidth]
{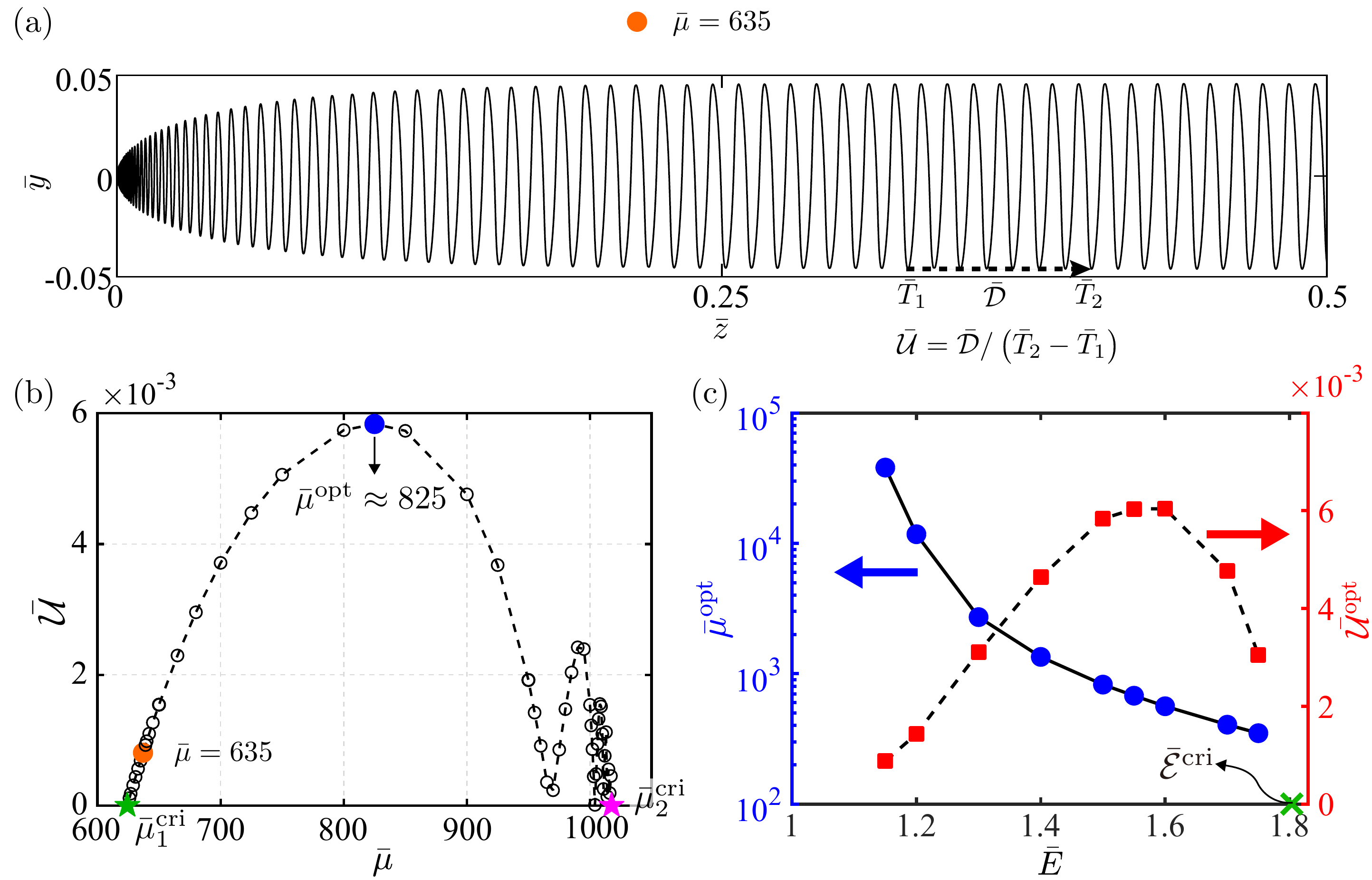}
\end{center}
\caption{(a) Trajectory of the particle centre for $\barmu=635$, when 
$\barE=1.5$. The dashed arrow indicates how the effective translational 
velocity  $\mU$
is quantified. (b) $\mU$ versus $\barmu \in \lp \barmucrione, \barmucritwo \rp$ 
when 
$\barE=1.5$, $\mU$ reaches an optimal value $\mUopt \approx 6\times 10^{-3}$ at 
$\barmu 
= 
\muopt \approx 825$. (c) The optimal EEV 
number 
$\muopt$ when the composite
object attains the maximum effective translational velocity 
$\mUopt$; $\muopt$ and $\mUopt$ are plotted versus the field 
strength $\barE$.}
\label{fig:opt_mu_and_spd}
\end{figure} 
 
Having demonstrated that the composite object is able to 
achieve propulsion by self-oscillatory undulation, we naturally 
examine its propulsive performance. Shown in figure~\ref{fig:opt_mu_and_spd}a, 
when the undulating swimmer reaches its time-periodic state,
its trajectory resembles a periodic wave propagating along a straight 
direction (dashed arrow). We thus define the effective translational velocity 
$\mU$ of the swimmer as the propagation speed of the wave, that is $\mU = 
\bar{\mathcal{D}}/\lp \bar{T}_2 - \bar{T}_1 \rp$. This effective velocity
$\mU$ exhibits a clear non-monotonic variation in $\barmu$; it reaches its 
maximum value at an optimal EEV number $\barmu = \muopt \approx 825$ and 
becomes zero when 
$\barmu \rightarrow \barmucrione$ and $\barmu \rightarrow \barmucritwo$.
Such a non-monotonic trend is expected, since when $\barmu$ is outside
the self-oscillating regime $[\barmucrione,\barmucritwo]$, the object
either remains stationary or spins steadily, resulting in no net 
locomotion. It is also worth noting that $\mU$ exhibits wavy variation near 
$\barmucritwo$. In this regime, the filament is so deflected
and the hydrodynamic interactions between the particle and filament 
can be reasonably strong due to the decreasing distance between them.
Since our simulations do not consider the hydrodynamic interactions, hence 
it is not self-consistent to interrogate the data in detail 
in this regime. 

Finally, we show in figure~\ref{fig:opt_mu_and_spd}c the 
dependence of the optimal swimming condition, $\muopt$ and $\mUopt$,  on the 
electric field strength $\barE$. The optimal EEV number $\muopt$ decreases
with $\barE$ monotonically; in contrast, the optimal velocity $\mUopt$ displays
a non-monotonic variation in $\barE$, reaching a maximum value of approximately
$6\times 10^{-3}$
at $\barE \approx 1.55 - 1.6$. This non-monotonic trend is not surprising.
In fact, self-oscillation of the composite object only emerges when $1 < \barE 
< \bar{\mathcal{E}}^{\mathrm{cri}}$, where $\bar{\mathcal{E}}^{\mathrm{cri}}$
represents the critical electric field above which the particle 
jointed with a rigid rod ($\barmu \rightarrow 0$) of the same length and 
slenderness will undergo the QR instability. Hence, when $\barE \geq 
\bar{\mathcal{E}}^{\mathrm{cri}}$, the composite object will spin steadily
but not self-propel regardless of the filament rigidity. 
On the other 
hand, when $\barE \leq 1$, the extra 
anchored filament will further stabilise the original QR particle, hence the 
composite object
will be stationary. We further note that the optimal translational velocity 
$\approx 6\times 10^{-3}$ is in the range $\lp1, 15\rp\times 10^{-3}$ of the 
dimensionless 
speed of a magnetically driven 
flexible artificial flagellum~\citep{dreyfus2005microscopic}.

\begin{figure}
\begin{center}
\hspace{0em}\includegraphics[width=0.6\textwidth]
{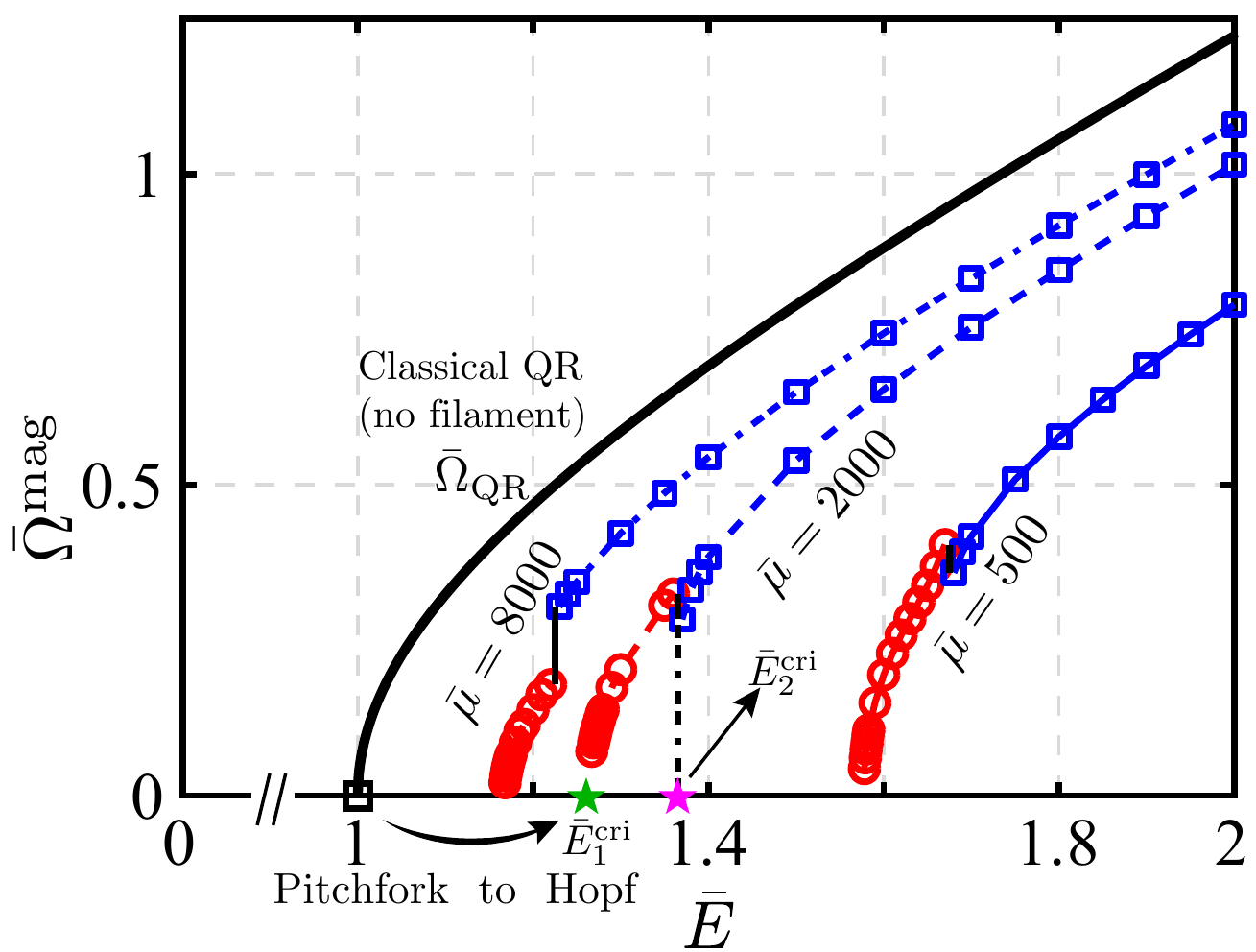}
\end{center}
\caption{Amplitude $\nonOmegaamp$ of the  
rotational velocity versus $\barE$ for three EEV numbers $\barmu=500$, 
$2000$ and $8000$. 
$\barEcrione$ (green star) and $\barEcritwo$ (magenta star) indicate
where the Hopf and secondary bifurcations occur, respectively.
The solid curve corresponds to the original QR rotational velocity, 
$\bar{\Omega}_{\mathrm{QR}}$ (see  
equation~(\ref{eq:non_omega_QR})) and the hollow square denotes $\barE=1$, 
the occurrence of the pitchfork bifurcation.}
\label{fig:omg_v_E_varyMu_a0.3}
\end{figure} 

By analogy to the results in figure~\ref{fig:omg_mu_alpha03_varyE}a-c, we show 
in figure~\ref{fig:omg_v_E_varyMu_a0.3}
$\nonOmegaamp$
versus $\barE$ as the bifurcation parameter for three EEV numbers $\barmu=500$, 
$2000$
and $8000$. A similar bifurcation diagram is identified: increasing $\barE$ 
from zero, the stationary fixed point solution transits to a limit-cycle 
solution through a supercritical Hopf bifurcation at $\barEcrione$ (green 
star); 
that solution then jumps to a second fixed point solution (steady spinning) via 
a secondary bifurcation at $\barEcritwo$ (magenta star).
The original QR instability emerges at $\barE=1$ (hollow square) through a 
supercritical pitchfork 
bifurcation~\citep{turcu1987electric,
peters2005experimental,das2013electrohydrodynamic}.
The filament manages to transform that bifurcation for an individual particle 
into a 
corresponding Hopf bifurcation leading to self-oscillation.
It is not surprising that by increasing $\barmu$, the variation of 
$\nonOmegaamp$ for the composite object tends to recover that of the original 
QR corresponding to $\barmu \rightarrow \infty$.

\begin{figure}
\begin{center}
\hspace{0em}\includegraphics[width=1\textwidth]
{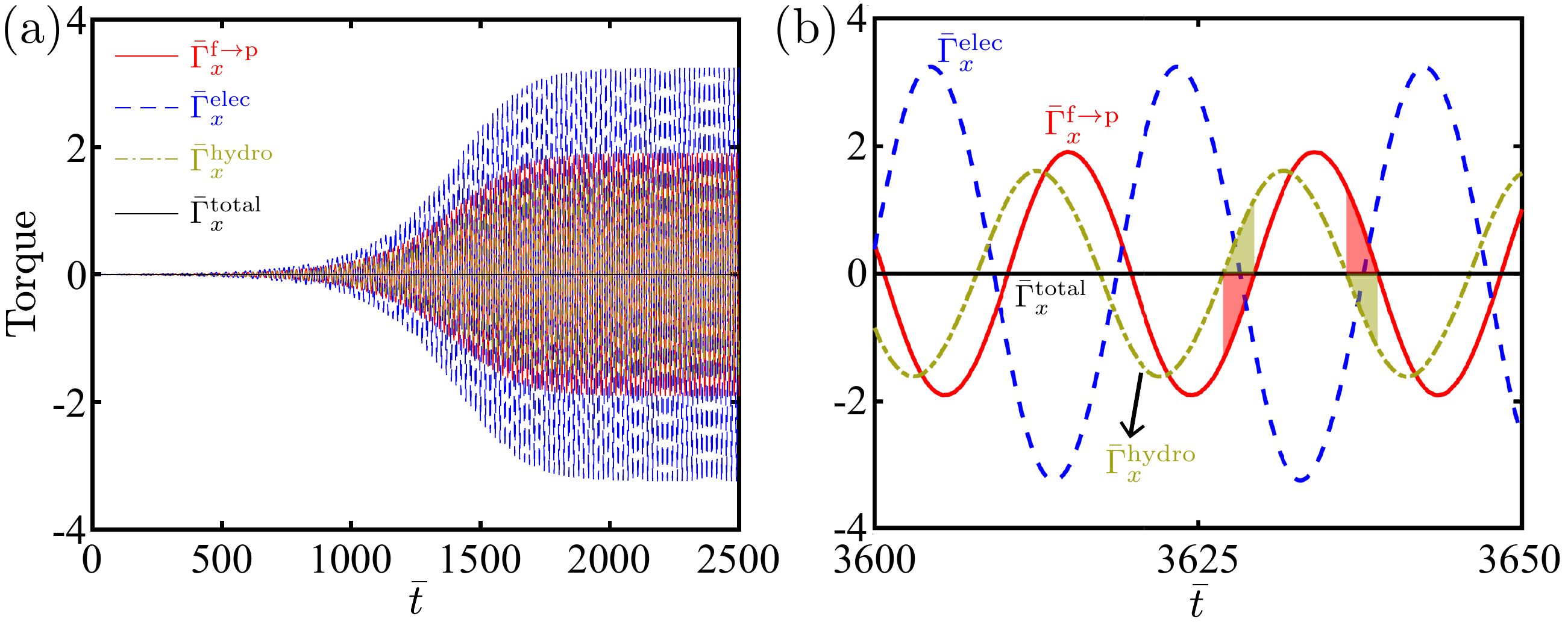}
\end{center}
\caption{(a) Time evolution of the elastic 
$\bar{\Gamma}_{x}^{\mathrm{f} \rightarrow \mathrm{p}}$ (solid), electric 
$\bar{\Gamma}_{x}^{\mathrm{elec}}$ (dashed), hydrodynamic 
$\bar{\Gamma}_{x}^{\mathrm{hydro}}$ (dot-dashed) and total 
$\bar{\Gamma}_{x}^{\mathrm{total}}$ (straight solid) torque ($x$-component) on 
the particle 
with respect to its centre, where $\barmu=635$ and $\barE=1.5$. (b) 
Close-up view of the time-periodic state. Shaded regions indicate when 
the elastic $\bar{\Gamma}_{x}^{\mathrm{f} \rightarrow \mathrm{p}}$ and 
hydrodynamic 
$\bar{\Gamma}_{x}^{\mathrm{hydro}}$ torques have opposite signs.}
\label{fig:torque_omg}
\end{figure}
It is evident that the elastic torque  $\nonbTfp$ plays 
an important role in 
the torque balance. We examine 
the time evolution of the $x$-component of the torques, namely the elastic 
$\bar{\Gamma}_{x}^{\mathrm{f} 
\rightarrow \mathrm{p}}$, hydrodynamic $\bar{\Gamma}_{x}^{\mathrm{hydro}}$ and 
electric 
$\bar{\Gamma}_{x}^{\mathrm{hydro}}$ torques in 
figure~\ref{fig:torque_omg} when 
$\barmu=635$ and 
$\barE=1.5$. The sum of the torques $\bar{\Gamma}_{x}^{\mathrm{total}} = 
 \bar{\Gamma}_{x}^{\mathrm{f} \rightarrow \mathrm{p}} + 
\bar{\Gamma}_{x}^{\mathrm{hydro}} + \bar{\Gamma}_{x}^{\mathrm{elec}}= 0$ 
implies 
that the torque balance is well satisfied numerically. Similar to the evolution 
of the
rotational velocity, the torques exhibit exponential growth in
the initial phase before approaching a time-periodic state. The torque balance 
in 
this state is further scrutinised in figure~\ref{fig:torque_omg}b. 
Realising the negative relation between $\nTvis_x$ and $\nonOmega$, 
we notice that $\nTfp_x$ and $\nonOmega$ have the same sign in the two 
highlighted periods emphasising when the elastic $\nTfp_x$ and $\nTvis_x$ 
hydrodynamic torque contributions have opposite signs. The in-phase behaviour 
of 
$\nTfp_x$ and $\nonOmega$ is a clear signature of negative damping, or 
positive feedback that triggers the linear instability of
self-oscillation~\citep{jenkins2013self}.

\section{Linear stability analysis}\label{sec:linear}
\subsection{Linearisation about the stationary equilibrium 
state}\label{sec:linear_qc}
We perform LSA about the stationary equilibrium state
of the composite particle when the filament is undeformed. In this 
section, we drop the bars for all of the dimensionless unknown variables 
(those 
over dimensionless parameters remain),  unless 
otherwise specified.
We linearise the governing equations of the particle orientation $\theta$, and
the dipole components $[\mP_Q, \mP_3]$. By incorporating into the LSA a 
theoretical 
model of the elasto-viscous response of the filament, we  do not 
linearise the equations for the filament position $\br(s)$ and tension $T(s)$
as conducted in \citet{guglielmini2012buckling}. 

The state variables $[\theta, \mP_Q, \mP_3]$ are decomposed into a
base (equilibrium) state  
$[\bstheta, 
\bsP_Q, \bsP_3]$
 and a perturbation state $[\ptheta, \pPQ, \pPT]$, which satisfy
  \begin{subequations} \label{eq:linear-decomp}
 \begin{align}
  \theta & = \bstheta + \ptheta, \\
 \mP_Q    & = \bsP_Q + \pPQ, \\
 \mP_3  & = \bsP_3 + \pPT.
 \end{align}
 \end{subequations}
The perturbation-state variables $[\ptheta,\pPQ, \pPT]$ are assumed to be 
infinitesimal in LSA.

By substituting  $\bTfp = \mathbf{0}$ and $\frac{\partial }{\partial t}=0$ into
equations~(\ref{eq:non-theta}), (\ref{eq:non-PQ}) and
\ref{eq:non-P3}, we obtain the base-state dipoles 
\begin{subequations}\label{eq:base-dipole}
\begin{align}
 \bsP_Q &= -\kappa \bareta \barE \sin{\bstheta},\\
 \bsP_3 &= -\kappa \bareta \barE \cos{\bstheta}.
\end{align} 
\end{subequations}
By substituting equations~(\ref{eq:linear-decomp}) and (\ref{eq:base-dipole})
into equations~(\ref{eq:non-theta}), (\ref{eq:non-PQ}) and
(\ref{eq:non-P3}), and assuming small $\ptheta$,
we derive the governing equations
for the perturbation-state variables $[\ptheta,\pPQ, \pPT]$,
\begin{subequations}\label{eq:pert_linear}
\begin{align}
\frac{\partial \pPQ}{\partial t} &= -\kappa \lp \pPQ + \kappa \bareta 
\barE 
\ptheta 
\cos{\bstheta} \rp,\\ 
\frac{\partial \pPT}{\partial t} &= -\kappa \lp \pPT - \kappa \bareta
\barE 
\ptheta 
\sin{\bstheta} \rp,\\ 
\frac{\partial \ptheta}{\partial t} &= \frac{1}{\bareta}\left[\Tfp_N +
 \barE \lp \pPQ\cos{\bstheta}-\pPT\sin{\bstheta} + \kappa\bareta 
\barE \ptheta 
\rp 
\right].\label{eq:pert_theta}
\end{align}
\end{subequations}

Adopting the normal-mode approach, we assume that the perturbations 
vary 
exponentially in time with a complex rate 
$\sigma = \sigr + \ii \sigi$, so
$\left[\pPQ, \pPT, \ptheta \right] = \left[ \Phi, \Pi, \Theta\right] 
\exp{(\sigma t)}$. Consequently,  
equation~(\ref{eq:pert_linear}) can be reformulated to
\begin{align}\label{eq:linear_qr}
\Tfp_N =  \frac{\sigma \left[ \sigma - (\barE^2-1)\kappa 
\right]}{\sigma+\kappa} 
\Theta \bareta \exp{(\sigma t)}.
\end{align} 
We note that, for a vanishing elastic torque $\Tfp_N=0$ (no attached 
filament), equation~(\ref{eq:linear_qr}) is characterised by two 
roots $\sigma_1=0$ 
and $\sigma_2 = 
\kappa \lp \barE^2-1 \rp$, which
describe
the original QR instability; the first root represents the stationary 
state and the second indicates that the dimensionless threshold electrical
field (scaled by $\Ecri$) required to trigger instability is $\barE=1$. 
Note 
that 
$\Ecri$ 
in equation~(\ref{eq:Ecri}) 
 is originally derived by balancing the electric and hydrodynamic
torque~\citep{jones1984quincke} instead of conducting LSA (see 
appendix~\ref{sec:append} for details). The two predictions
exactly agree with each other.

\subsection{Elastohydrodynamic model}\label{sec:eh_model}
Since the elastohydrodynamic
equations are not linearised, we thus derive a theoretical expression 
for $\Tfp_N (t)$ for the dispersion relation, 
equation~(\ref{eq:linear_qr}).
\begin{figure}
\begin{center}
\hspace{0em}\includegraphics[width=1\textwidth]
{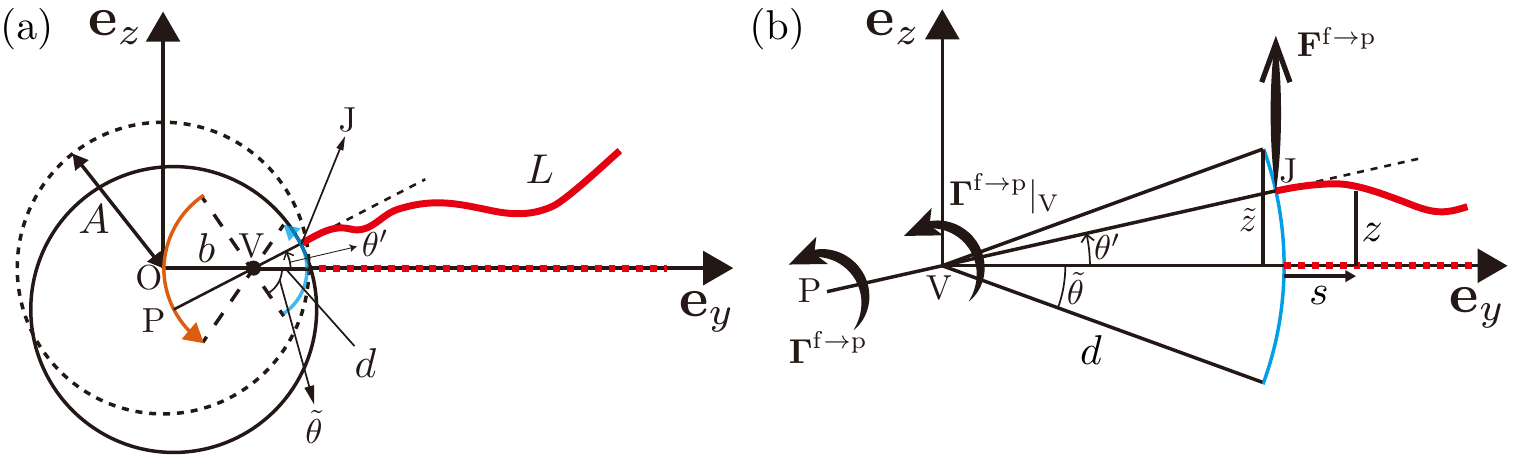}
\end{center}
\caption{(a) Schematic of the model problem: a composite object of a sphere and 
a filament undergoes a rotational oscillation. The particle centre P and joint 
(filament base) J rotate periodically along 
circular arcs of radius $b$ and $d=A-b$, respectively. V denotes their 
common pivot point. (b) Zoom-in 
on the circular arc trajectory of the joint, showing the difference between the
torque $\bTfp|_{V}$ with respect to the pivot V and $\bTfp$ to 
the particle centre P.}
\label{fig:force-torque-sketch}
\end{figure}
We find $\Tfp_N (t)$  by solving a separate
elastohydrodynamic problem of the composite object undergoing a prescribed
rotational oscillation characterised by $\ptheta = \pthemag (t)\exp{\lp \ii \si 
t\rp}$, where $\pthemag (t) = \Theta \exp{\lp \sr t 
\rp}$ indicates the angular 
oscillation amplitude. We do not consider the object's translation near the 
onset of instability since any translation is 
negligible due to the small-amplitude oscillation.
To simplify the algebra in the next steps,
we set $\bstheta = \pi/2$ without loss of
generality as
shown in figure~\ref{fig:force-torque-sketch}, where the rest 
configuration (dashed curves) corresponds to
when the particle centre P coincides with the origin O and the undeformed
filament is aligned in the $\be_y$ direction.  The 
rotational oscillation is executed 
about a pivot V that lies  away from the origin by a dimensional
distance $b$ on the $y$-axis, 
where $\beta = b/L$; the dimensional distance between 
V and J is $d = A-b$, so similarly 
\begin{align}\label{eq:delta}
\delta 
= d/L = 
\alpha - \beta.
\end{align}
The particle centre P (resp. filament base J) follows a 
trajectory
of a circular arc that is centred at V and of radius $b$ (resp. $d$); both 
trajectories are symmetric about the $y$-axis. Note that $\beta$ is an unknown 
that is to be 
determined.

Near the onset of instability, the amplitude $\pthemag (t)$ 
varies much more slowly than the oscillation of $\ptheta$, \textit{viz.} $\sr 
\ll 
\si$. This allows us to assume that the amplitude 
$\pthemag = \Theta \exp{(\sr t)}$ is quasi-steady, namely, $\ptheta$ at a 
particular time $\tp$ can be approximated by
\begin{align}
 \ptheta = \Theta \exp{\lp\sr \tp\rp} \exp{\lp \ii \si t \rp},
\end{align}
as an instantaneous configuration of a periodic signal with a prescribed 
amplitude 
$\Theta \exp{\lp\sr \tp\rp}$ and frequency $\si$. This setup resembles 
the theoretical framework 
developed to address the so-called elastohydrodynamic problem 
II~\citep{wiggins1998flexive,wiggins1998trapping} of a filament with one of its 
ends undergoing straight, oscillatory 
translation. We adapt that framework for our configuration, whereas the
filament 
end oscillates on a circular arc instead of on a straight path, as shown in
figure~\ref{fig:force-torque-sketch}b. Because 
the filament 
undergoes small-amplitude deformation, $|z_y| \ll 1$ and its tangent vector 
$\br_s \approxeq \be_y$. We also assume $T(s) \equiv 0$. The 
position $\br(t, s)$ of the filament centreline is $\br(s) = 
(\alpha 
+ s) \be_y + z(t, s) \be_z$. 
The horizontal displacement of the filament base is of order 
$\mathcal{O}(\pthemag^2)$ and can be neglected because $|\ptheta| 
\leq |\pthemag| \ll 1$.  The base's  
vertical oscillation is prescribed as
\begin{align}\label{eq:z_s0}
 z(t)|_{s=0} & = \delta \sin{\ptheta}  \approxeq \delta \ptheta = \delta 
\pthemag \exp{(\ii \si t)},
\end{align}
where $\delta \pthemag$ represents the oscillation amplitude.
Following \citet{wiggins1998flexive} and \citet{wiggins1998trapping}, the
deflection of the filament is expressed by
\begin{align}\label{eq:z_s}
 z (s) = \delta \pthemag \exp{(\ii \si t) h(s,\mL)},
\end{align}
where 
\begin{align}\label{eq:mL}
\mL^4 = \frac{\barmu \si}{-1-2\log{\esl}}
\end{align}
and $h$ is a sum of four
solutions
\begin{align}
 h(s,\mL) = c_1 \xi^{\ii s} + c_2 \xi^{-s} + c_3 \xi^{-\ii s} + c_4 \xi^{s},
\end{align}
with
\begin{subequations}\label{eq:z0_xi}
 \begin{align}
 z_0 & = \exp{(-\ii \pi/8)}, \label{eq:z0} \\
\xi & = \exp{(z_0 \mL)}. \label{eq:xi} 
 \end{align}
\end{subequations}
 The four coefficients $c_i$ need to be determined by the BCs at the filament 
ends.
In contrast to \citet{wiggins1998flexive} and \citet{wiggins1998trapping} 
treating $z(s)$ as 
a real variable, we consider a complex $z(s)$. This allows us to obtain
the complex torque consistent with
the complex nature of the torque balance, equation~(\ref{eq:linear_qr}).

The BCs for $h(s)$ at the free end $s=1$ are $h_{ss}=h_{sss}=0$. At the clamped 
end 
$s=0$, $h=1$ as a Dirichlet BC corresponding to the prescribed displacement; 
the 
other
BC is more subtle. Because the filament 
orientation is orthogonal
to the circular arc (see figure~\ref{fig:force-torque-sketch}b),  we have
\begin{align}\label{eq:bc_zs}
 z_s & = \sin{\ptheta} \approxeq \ptheta =\pthemag \exp{(\ii \si t)}.
\end{align}
By substituting equation~(\ref{eq:z_s}) into equation~(\ref{eq:bc_zs}), we 
obtain 
the BC
\begin{align}
 h_s|_{s=0} & = 1/\delta,
\end{align}
where $\delta$ is defined in equation~(\ref{eq:delta}).
Knowing all the BCs of $h(s)$, we compute the four coefficients
\begin{subequations}
 \begin{align}
  c_1 & = \frac{\left(1+\ii\right) \left[\left((1-\ii) \xi 
^{1+\ii}-\ii \xi ^2+1\right) \delta \log \xi -(1+\ii) \xi ^{1+\ii}-\ii \xi 
^2-1\right]}{2  \Lambda
\delta \log \xi }, \\
  c_2 & = \frac{\left(1+\ii \right) \xi  \left[ \left(-\ii \xi 
^{1+2 \ii}+(1-\ii) \xi ^\ii+\xi \right) \delta \log \xi-\xi ^{1+2 
\ii}+(-1+\ii) \xi ^\ii+\ii \xi \right]}{2 \Lambda \delta \log \xi },\\
  c_3 & = \frac{\left( 1 + \ii \right) \xi ^\ii \left[\left(\xi 
^{2+\ii}-\ii \xi ^\ii+(1-\ii) \xi \right) \delta \log \xi+\xi ^{2+\ii}+\ii \xi 
^\ii+(1+\ii) \xi \right]}{2 \Lambda \delta \log\xi},\\
  c_4 & = \frac{(1+\ii) \left(\xi ^{2 \ii}+(1-\ii) \xi ^{1+\ii}-\ii\right) 
\delta \log (\xi )+(1-\ii) \xi ^{2 \ii}+2 \xi ^{1+\ii}+1+\ii}{2 \Lambda 
\delta \log \xi },
 \end{align}
\end{subequations}
where $\Lambda=\xi ^{2 \ii}+4 \xi ^{1+\ii}+\xi ^{2+2 \ii}+\xi 
^2+1$. Considering the small-amplitude deformation, the total
force $\bF$ exerted by the filament on the clamped end 
is along the vertical $\be_z$ direction. The torque 
$\bTfp|_{V}$
with respect to the pivot V and $\bTfp$ with respect to the particle 
centre P 
are along the $\be_x$ direction, so that the corresponding components 
of the force and torques are 
\begin{subequations} \label{eq:force-torque-full}
\begin{align}
 \Ffp_z & = \pthemag \exp{\lp \ii \si t \rp} \frac{\log^2{\xi}\left[(1+\ii) 
\Lambda_1 \delta \log\xi - \ii \Lambda_2 \right]}{\Lambda}, 
\label{eq:force-full}\\
\Tfp_x|_{V} & = \pthemag \exp{\lp \ii 
\si t \rp} \frac{\log \xi  \left[(1+\ii) 
 \delta^2 
\Lambda_1 \log^2\xi  - 2\ii \delta  \Lambda_2 \log\xi  +(-1-\ii) 
\Lambda_3 \right]}{\Lambda},\\
\Tfp_x & = \Tfp_x|_{V} + \lp \alpha - \delta \rp \Ffp_z \nonumber \\
& = \pthemag \exp{\lp \ii 
\si t \rp} \frac{\log \xi  \left[(1+\ii) 
\alpha  \delta 
\Lambda_1 \log^2\xi  -\ii (\alpha + \delta ) \Lambda_2 \log\xi  +(-1-\ii) 
\Lambda_3 \right]}{\Lambda},
\end{align} 
\end{subequations}
where 
\begin{subequations}
 \begin{align}
  \Lambda_1 & = -\xi ^{2 \ii}-\ii \xi ^{2+2 \ii}+\xi ^2+\ii, \\
  \Lambda_2 & = \left(-1+\xi ^{2 \ii}\right) \left(\xi ^2-1\right),\\
  \Lambda_3 & = \ii \xi ^{2 \ii}+\xi ^{2+2 \ii}-\ii \xi 
^2-1.
 \end{align}
\end{subequations}

\begin{figure}
\begin{center}
\hspace{0em}\includegraphics[width=1\textwidth]
{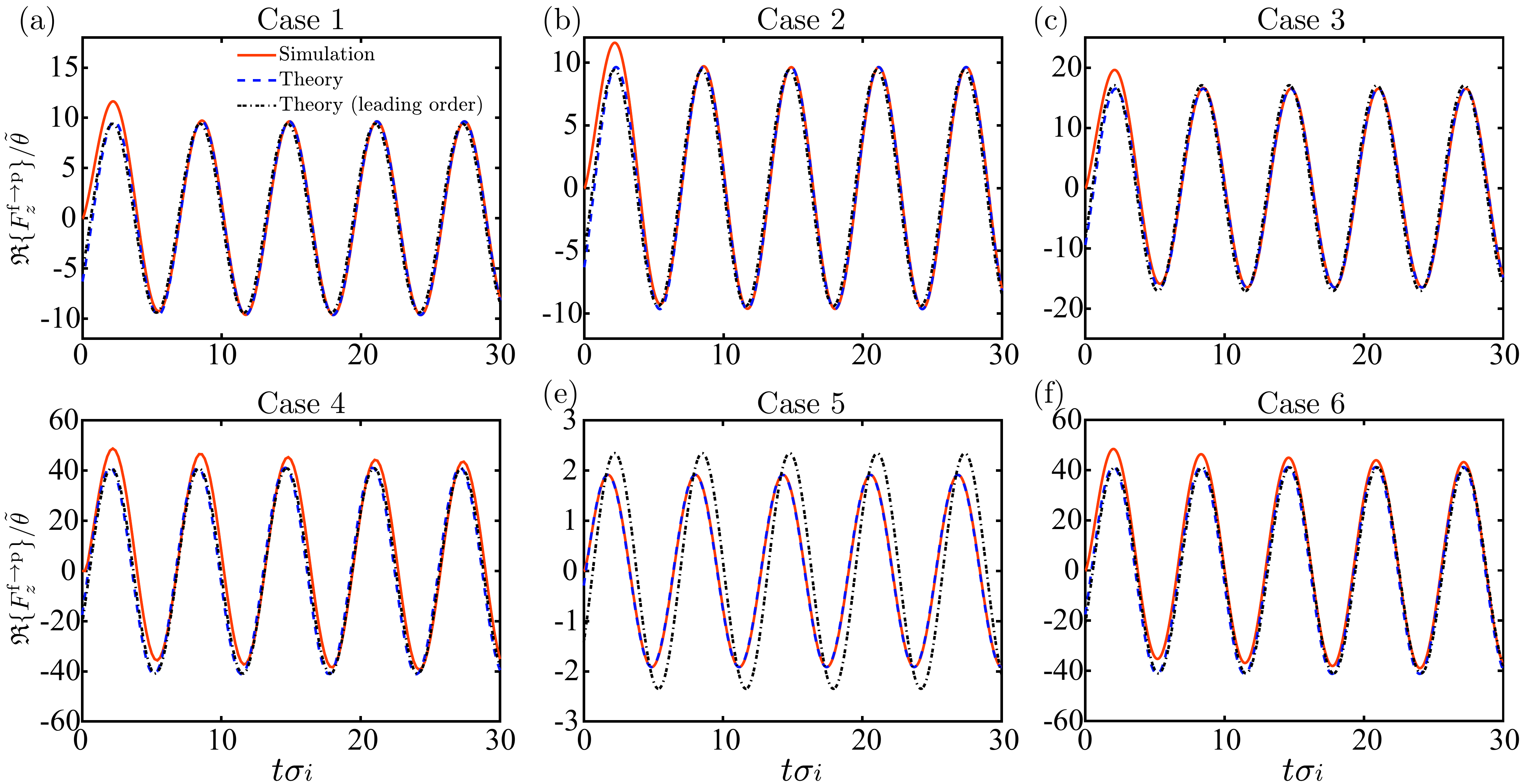}
\end{center}
\caption{Comparison between the theoretical force $\Ffp_z$ (dashed curves), 
its leading-order approximation $\sFfp_z$ (dot-dashed curves) and the numerical 
results (solid curves).}
\label{fig:force-valid}
\end{figure}

\begin{figure}
\begin{center}
\hspace{0em}\includegraphics[width=1\textwidth]
{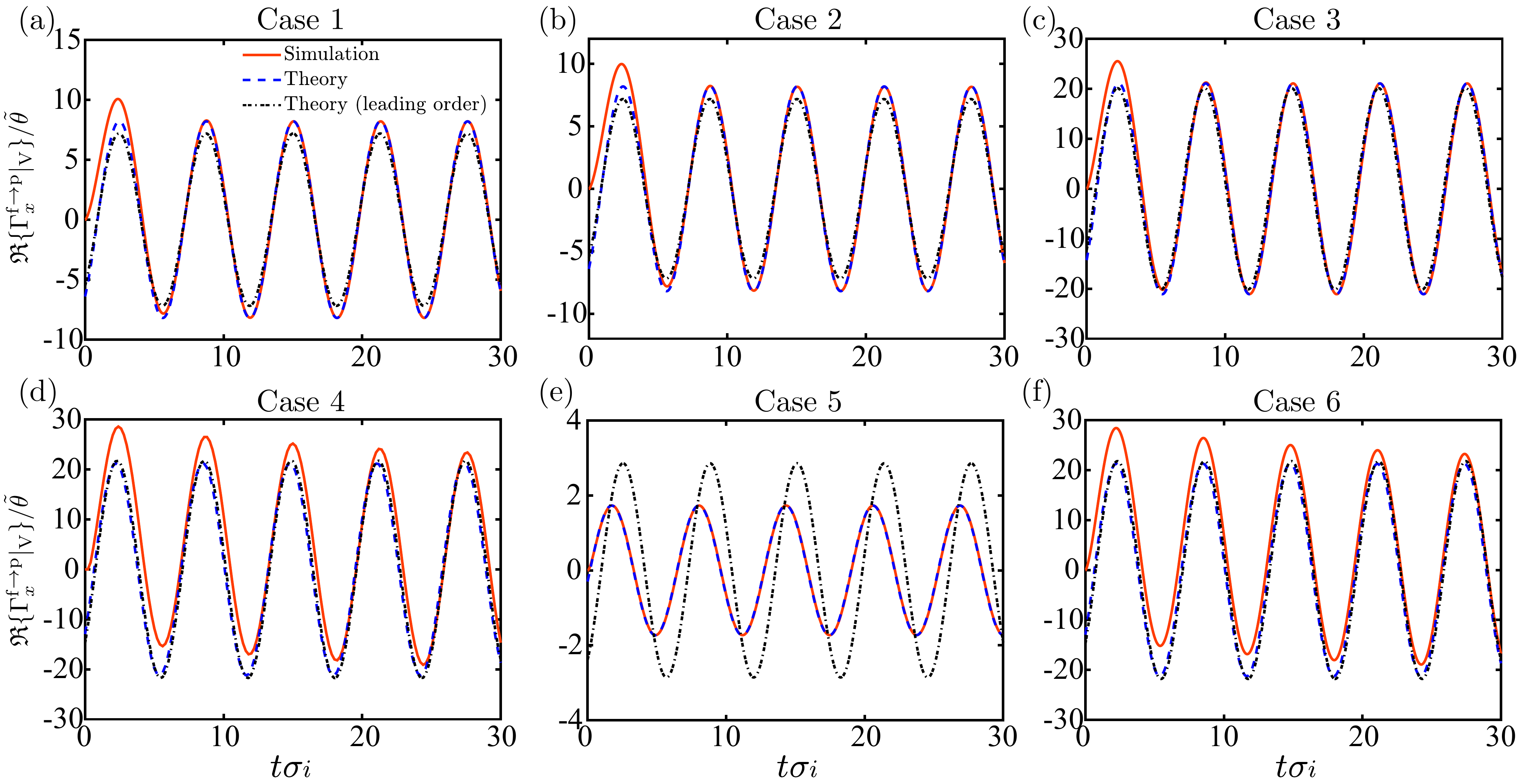}
\end{center}
\caption{Comparison between the theoretical torque $\Tfp_x|_{\mathrm{V}}$ 
(dashed curves), 
its leading-order approximation $\sTfp_x|_{\mathrm{V}}$ (dot-dashed curves) and 
the numerical 
results (solid curves).}
\label{fig:torque-valid}
\end{figure}

Now, let us examine the denominator, $\Lambda$, of 
equation~(\ref{eq:force-torque-full}) whose five terms are
in the form of $\xi^{q_k}$ ($k=1 ... 5$), where $[q_1, q_2, q_3, 
q_4, q_5]=[2\ii, 1+\ii, 2+2\ii, 2, 0]$.
Using equation~(\ref{eq:z0_xi}), we  express $\xi^{q_k}$ as 
\begin{align}
 \xi^{q_k} = \left[\exp{(z_0 \mL)}\right]^{q_k} = \zeta^\mL_k,
\end{align}
where $\zeta_k = \exp{(z_0 q_k)}$ are
\begin{align}
 \zeta_1 & = -0.59 + 2.09\ii,\nonumber \\
 \zeta_2 & = 3.17 + 1.9\ii, \nonumber \\
 \zeta_3 & = 6.4 + 12.05 \ii, \nonumber \\
 \zeta_4 & = 4.57 - 4.4\ii, \nonumber \\
 \zeta_5 & = 1.
\end{align}
We observe that the third term $\zeta_3^{\mL}$ is larger than the 
rest in magnitude when
$\mL \geq 1$, dominating the second largest term by one order 
when $\mL \geq 3$. Let us assume 
$\mL\geq3$ a priori, so that we can then approximate $\Lambda$ by
$\zeta_3^{\mL}$ in equation~(\ref{eq:force-torque-full}). By further extracting 
the 
leading-order terms of $\Lambda_1/\Lambda$,
$\Lambda_2/\Lambda$ and $\Lambda_3/\Lambda$,  
we attain a simplified, leading-order expression for the force and torque 
(denoted by 
$\;\tilde\;\;$)
\begin{subequations}\label{eq:force-torque-simple}
\begin{align}
\sFfp_z & = \pthemag \exp{\lp \ii \si t \rp} \log^2{\xi}\left[(1-\ii)  \delta 
\log\xi-\ii \right], \label{eq:force-simple}\\
\sTfp_x|_{\mathrm{V}} & =\pthemag \exp{\lp \ii 
\si t \rp} \log \xi  \left[(1-\ii) 
\delta^2 
\log^2\xi -2\ii \delta  \log\xi -1-\ii \right], \\
\sTfp_x & =\pthemag \exp{\lp \ii 
\si t \rp} \log \xi  \left[(1-\ii) 
\alpha  \delta 
\log^2\xi -\ii ( \alpha 
+\delta ) \log\xi -1-\ii \right].\label{eq:torque_simple}
\end{align} 
\end{subequations}
The theoretical force $\Ffp_z$, torque $\Tfp_x|_{\mathrm{V}}$ 
and their leading-order counterparts $\sFfp_z$ and $\sTfp_x|_{\mathrm{V}}$
are validated against the numerical results
for six cases spanning a wide range of parameters 
relevant to our study (see table.~\ref{tab:sixcases}), where case 1 is the 
reference case and the other five
vary a single parameter compared to case 1.
Because the numerical force and torque are real quantities, the 
real parts of $\Ffp_z$ (dashed curve) given by equation~(\ref{eq:force-full}), 
and its
leading-order approximation $\sFfp_z$ (dot-dashed curve) by 
equation~(\ref{eq:force-simple}), are compared with the numerical data
(solid curve) in 
figure~\ref{fig:force-valid}. A similar comparison between the 
torques $\Tfp_x|_{\mathrm{V}}$ and $\sTfp_x|_{\mathrm{V}}$ is shown in 
figure~\ref{fig:torque-valid}.

\begin{table}
\center
\begin{tabular}{ccccc}
\hline
\multicolumn{1}{|l|}{} & \multicolumn{1}{c|}{$\pthemag$} & 
\multicolumn{1}{c|}{$\delta$} & \multicolumn{1}{c|}{$\si$} & 
\multicolumn{1}{c|}{$\barmu$} \\ \hline
\multicolumn{1}{|c|}{Case 1 (reference)} & 
\multicolumn{1}{c|}{$10^{-3}$} & 
\multicolumn{1}{c|}{$0.3$} & \multicolumn{1}{c|}{$0.2$} & 
\multicolumn{1}{c|}{$10^3$} \\ \hline
\multicolumn{1}{|c|}{Case 2} & \multicolumn{1}{c|}{$\mathbf{0.1}$} & 
\multicolumn{1}{c|}{$0.3$} & \multicolumn{1}{c|}{$0.2$} & 
\multicolumn{1}{c|}{$10^3$} \\ \hline
\multicolumn{1}{|c|}{Case 3} & \multicolumn{1}{c|}{$10^{-3}$} & 
\multicolumn{1}{c|}{$\mathbf{0.8}$} & \multicolumn{1}{c|}{$0.2$} & 
\multicolumn{1}{c|}{$10^3$} \\ \hline
\multicolumn{1}{|c|}{Case 4} & \multicolumn{1}{c|}{$10^{-3}$} & 
\multicolumn{1}{c|}{$0.3$} & \multicolumn{1}{c|}{$\mathbf{2}$} & 
\multicolumn{1}{c|}{$10^3$} \\ \hline
\multicolumn{1}{|c|}{Case 5} & \multicolumn{1}{c|}{$10^{-3}$} & 
\multicolumn{1}{c|}{$0.3$} & \multicolumn{1}{c|}{$0.2$} & 
\multicolumn{1}{c|}{$\mathbf{10^2}$} \\ \hline
\multicolumn{1}{|c|}{Case 6} & \multicolumn{1}{c|}{$10^{-3}$} & 
\multicolumn{1}{c|}{$0.3$} & \multicolumn{1}{c|}{$0.2$} & 
\multicolumn{1}{c|}{$\mathbf{10^4}$} \\ \hline
                       &                       &                       &        
 
              &                      
\end{tabular}
\caption{Parameters for the six cases chosen to validate numerical 
results against the theoretical force $\Ffp_z$, torque 
$\Tfp_x|_{\mathrm{V}}$ 
and their leading-order counterparts. Bold entries 
indicate the difference with the reference, case $1$.}
\label{tab:sixcases}
\end{table}

We observe that the force  $\Ffp_z$ and torque $\Tfp_x|_{\mathrm{V}}$
and their leading-order values
 agree 
with the numerical results quantitatively in all the cases except for
 case $5$, where the leading-order results deviate a little from the
full expression and numerical results. This disagreement results
from the violation of the assumption $\mL \geq 3$ used to derive the 
leading-order expression, where $\mL \approx 1.25$ for case $5$. This
also implies that the leading-order predictions become less accurate at
small $\barmu$ values.

For the validation purpose, $\delta=\alpha-\beta$ can be prescribed.
However, for the model,
 $\delta$ needs to be determined using the force-free condition on the 
particle. The particle follows a circular arc on the 
other 
side of the pivot V, the $z$-component of the hydrodynamic force on the 
particle approximated by
the Stokes's law is
\begin{align}\label{eq:force_par}
F_z^{\mathrm{h}\rightarrow \mathrm{p}} = \frac{3\ii}{4}\barmu \alpha\beta\si 
\pthemag \exp{\lp \ii \si t \rp}.
\end{align}
Substituting equation~(\ref{eq:force-simple}) and (\ref{eq:force_par})  into 
$\sFfp_z + 
F_z^{\mathrm{f}\rightarrow 
\mathrm{p}}=0$, we obtain
\begin{align}
\beta  
& = \frac{4\log^2{\xi}\left[ 1 + \alpha \lp \ii + 1 \rp \log{\xi} \right]
}{3 \alpha \barmu \si+ 4 \lp \ii +1 \rp \log^3{\xi}}.
\end{align}
Using the leading-order torque $\sTfp_x$ 
equation~(\ref{eq:torque_simple}),
as the left-hand side torque of equation~(\ref{eq:linear_qr}) (note that the 
nodal line direction 
$\bN=\be_x$ when the orientation $\ortp$ is restricted to
the 
$yz$-plane), we 
obtain the governing equation for the transformed growth rate $\hsig = 
\barmu \sigma$,
\begin{align}\label{eq:hsig}
 \alpha^3 \frac{\hsig\left[ \hsig - \lp \barE^2-1 \rp \kappa \barmu \right] 
}{\hsig 
+ \kappa \barmu} + 
\log \xi  
\left[(\ii-1) 
\alpha  (\alpha -\beta ) 
\log^2\xi + \ii (2 \alpha 
-\beta ) \log\xi +1+\ii \right]=0,
\end{align}
where
\begin{align}\label{eq:beta}
 \beta  
& = \frac{4\log^2{\xi}\left[ 1 + \alpha \lp \ii + 1 \rp \log{\xi} \right]
}{3 \alpha \hsig_i+ 4 \lp \ii +1 \rp \log^3{\xi}},
\end{align}
where $\log{\xi}$ can be written as
\begin{align}\label{eq:log_xi}
 \log{\xi} = z_0 \mL = z_0 \lp \frac{\hsig_i}{-1-2\log{\esl}} \rp^{1/4}.
\end{align}

\subsection{Complex growth rates and onset of 
instability}\label{sec:lsa2}
We solve equation~(\ref{eq:hsig}) to obtain the transformed growth rate 
$\hsig$, which facilitates a theoretical prediction of the onset of 
self-oscillatory instability. 
The growth rate $\hsig =\hsig_r + \ii \hsig_i$ depends on 
$\alpha$, $\esl$, $\kappa$, $\barmu$ and $E$, where  we have fixed 
$\esl$ and $\kappa$. 
By writing $\hsig_i = W^4$ and substituting it into equation~(\ref{eq:mL}), we 
obtain $\mL = W/(-1-2\log{\esl})^{1/4}$. Here, $\mL$ is a positive real 
number, so is $W$. By substituting equations~(\ref{eq:beta})
and (\ref{eq:log_xi}) into equation~(\ref{eq:hsig}), we derive a system of
two-dimensional, nonlinear polynomial equations for $\hsig_r$ and $W$ (see
appendix~\ref{sec:polynomial}) and obtain its roots by employing 
the python driver phcpy~\citep{verschelde2013modernizing,otto2019solving} of a 
general-purpose 
solver PHCpack~\citep{verschelde1997phcpack} for polynomial 
systems. Because $\hsig=\barmu\sigma$, we obtain the real part $\sigma_r = 
\hsig_r/\barmu$ and imaginary part $\sigma_i = W^4/\barmu$ of the complex 
growth rate $\sigma$.
\begin{figure}
\begin{center}
\hspace{0em}\includegraphics[width=1\textwidth]
{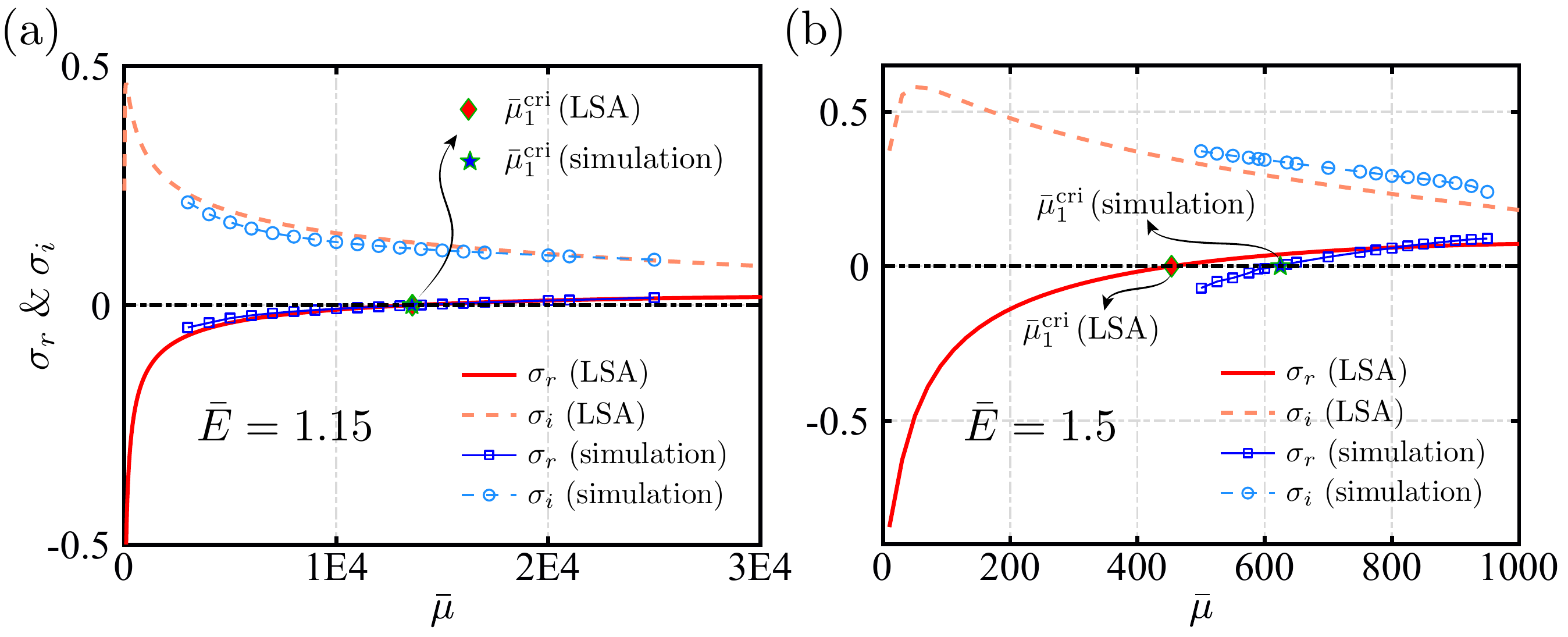}
\end{center}
\caption{The real $\sigr$ and imaginary $\sigi$ part of 
the complex growth rate $\sigma = \sigr + \ii \sigi$ versus $\barmu$ for two 
electric fields (a) $\barE=1.15$ and (b) $\barE=1.5$, where
the size ratio $\alpha=0.3$. Theoretical (LSA) and numerical predictions are 
denoted by red curves and blue symbols, respectively. The intersection of 
$\sigr(\barmu)$
with $\sigma=0$ gives the critical EEV value $\barmucrione$ (indicated by 
diamonds and pentagrams for theoretical and numerical results, respectively) 
corresponding to the onset of instability.}
\label{fig:sigma_v_mu}
\end{figure}

We show $\sigma_r$ and $\sigma_i$ as a function of $\barmu$ in 
figure~\ref{fig:sigma_v_mu} for two electric fields $\barE=1.15$ (a) and 
$\barE=1.5$ (b), where $\alpha=0.3$. In both cases, the imaginary part 
$\sigma_i \lp \barmu \rp > 0$ implying that the perturbation always 
decays/grows in an oscillatory manner.
In contrast, the real part $\sigma_r$ increases with $\barmu$ monotonically
from negative to positive values, indicating the critical condition 
$\sigma_r \lp  \barmucrione\rp = 0$  of the
self-oscillatory instability. When $\barmu$ is smaller/larger
than $\barmucrione$, the perturbation exhibits oscillatory decaying and growth. 
The LSA prediction of $\lp\sigma_r, \sigma_i\rp$ agrees quantitatively with the 
numerical counterpart for the $\barE=1.15$ case, and qualitatively for the 
$\barE=1.5$ case.

\begin{figure}
\begin{center}
\hspace{0em}\includegraphics[width=0.6\textwidth]
{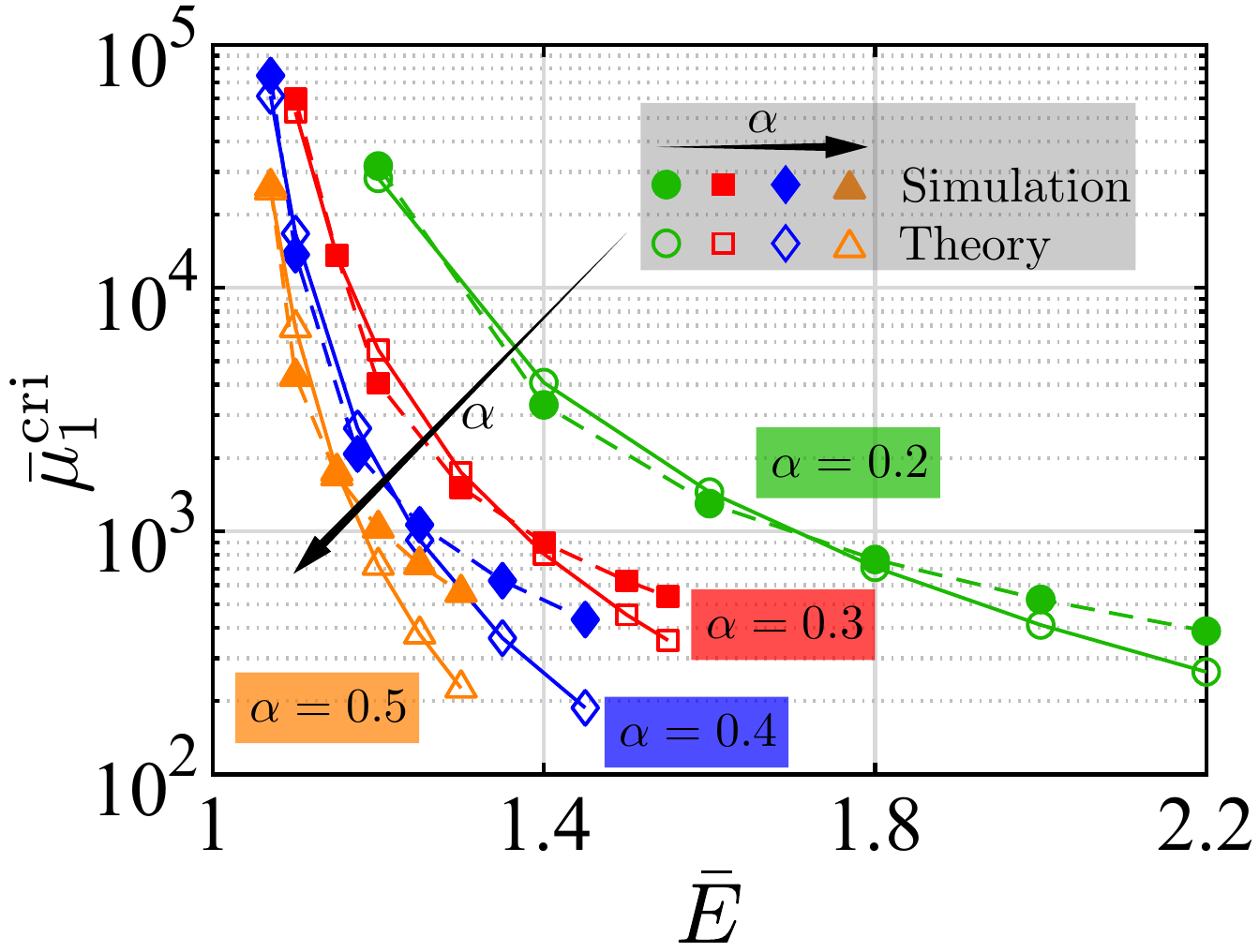}
\end{center}
\caption{The LSA (hollow symbols) and numerical (filled symbols) predictions of 
the critical EEV number $\barmucrione$ 
(versus $\barE$) at which instability occurs through a Hopf bifurcation, for 
size ratios 
$\alpha=0.2$ (circles), $0.3$ (squares), $0.4$ (diamonds) and $0.5$ 
(triangles). 
}
\label{fig:crimu_v_E}
\end{figure}

We adopt a bi-section method to determine $\barmucrione$ as a function of 
$\lp\barE,\alpha\rp$, as shown in figure~\ref{fig:crimu_v_E}.
For all $\alpha$ values, $\barmucrione$ decreases monotonically with $\barE$.
The theoretical and numerical predictions agree well with each other, 
especially in the high $\barmu$ regime. The agreement 
degenerates with decreasing $\barmu$. We infer that $\mL$ becomes 
smaller when 
$\barmu$ decreases, hence this disagreement is mostly attributed to violating 
the 
$\mL \geq 3$ assumption of the leading-order force/torque model
for the LSA.

\section{A minimal model to reproduce the EEH instability and 
self-oscillation}\label{sec:minimal}
\begin{figure}
\begin{center}
\hspace{0em}\includegraphics[width=0.75\textwidth]
{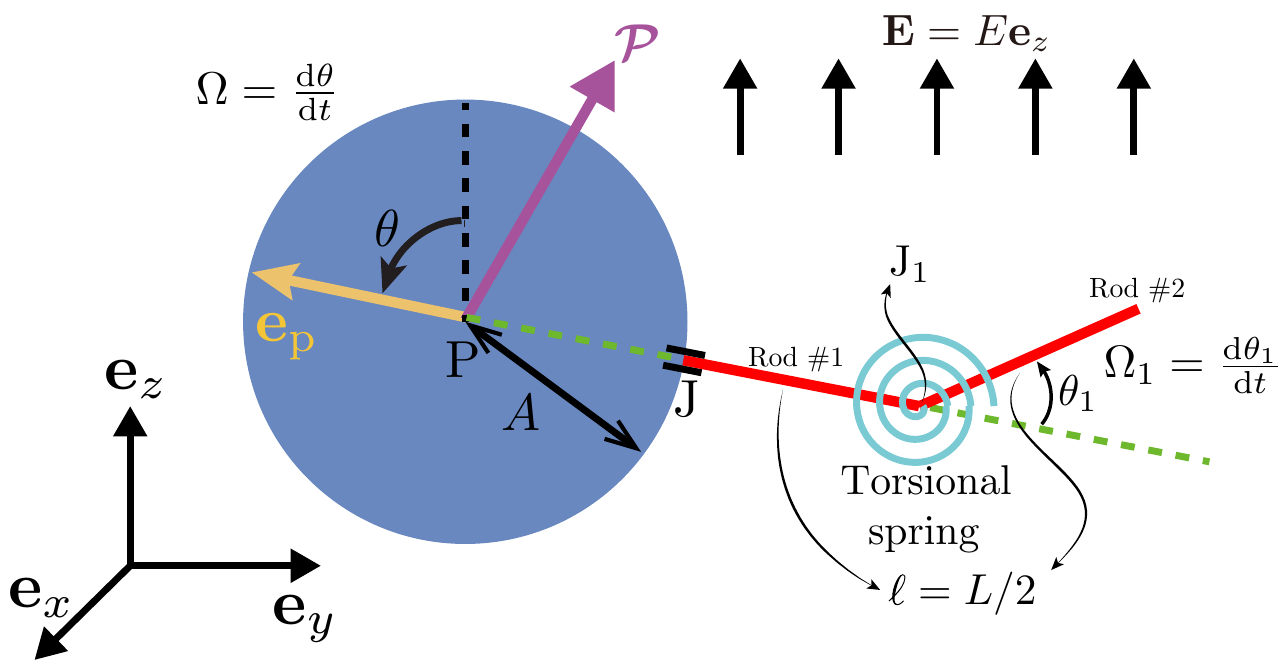}
\end{center}
\caption{Schematic of a minimal model to reproduce the EEH instability and 
self-oscillation: the elastic 
filament is represented by two rigid rods of the same length $\ell=L/2$ 
linked flexibly at $\Jone$ by a torsional spring of elastic modulus $K$.
Rod $\#1$ is rigidly anchored at J. A steady, uniform electrical field 
$\bE=E\be_z$ is applied.}
\label{fig:sketch-linker}
\end{figure}
To better unravel the physics underlying the EEH instability, we seek 
a minimal model  reproducing this instability and the corresponding 
self-oscillation. By analogy to the multi-linker 
models~\citep{de2017spontaneous,ling2018instability}, we replace 
the elastic filament by two rigid cylindrical rods numbered $\#1$ and $\#2$ of 
equal length $\ell =L/2$ and equal radius $a$ of their cross sections, which 
are linked at $\Jone$ by a torsional spring with an elastic module of $K$ (see 
figure.~\ref{fig:sketch-linker}). Rod $\#1$ is clamped at the sphere surface 
J, namely it always passes through the particle centre P, hence the 
displacement vectors $\overrightarrow{\mathrm{PJ}}$ and 
$\overrightarrow{\mathrm{P}\Jone}$ are 
opposite to the particle orientation $\ortp$. Rod $\#2$ is oriented 
with respect to $\overrightarrow{\mathrm{P}\Jone}$ by an angle $\theta_1$, 
which 
is zero 
when the composite system is at rest.

Similar to the original setup, we 
assume that the motion 
of particle and the rods are restricted to the $yz$-plane. Further, 
no hydrodynamic interactions between the particle and rods, or between the rods 
are considered. The system consists of six unknowns: the 
translational velocity components $U_y(t)$ and $U_z(t)$ of the particle, the 
rotational velocity component $\frac{\d \theta(t)}{\d t} = \Omega(t)$ of the 
particle and 
$\frac{\d \theta_1(t)}{\d t} = \Omega_1 (t)$ of rod $\#2$ with 
respect to rod $\#1$, and the polarisation vector components $\mP_Q(t)$ and 
$\mP_3(t)$. 
It is worth noting that compared to the classical QR particle, this minimal 
configuration only incorporates one extra degree of freedom,  
$\theta_1$, which indicates the deformation magnitude of the torsional spring.

We first derive the hydrodynamic force 
exerted on rod $\#1$ as
\begin{align}
 \mathbf{F}^{\mathrm{hydro}}_{1}  = & \frac{2 \pi \mu \ell  }{c} \left[2 
\theta_t  (2 A+\ell) \cos \theta+U_y \cos2 \theta +3 U_y+U_z \sin 2 
\theta \right]\be_y \nonumber \\
+ &\frac{2 \pi   \mu \ell}{c} \left[2 \theta_t  (2 A+\ell) \sin \theta+U_y 
\sin 2 \theta - U_z \cos{2\theta} +3 U_z\right]\be_z,
\end{align}
and the torque about the particle centre P 
\begin{align}
 \boldsymbol{\Gamma}^{\mathrm{hydro}}_{1}|_{\mathrm{P}} = \frac{4 \pi   \mu 
\ell }{3 c} \left[ 2 \theta_t  \left(3 A^2+3 A \ell+\ell^2\right)+3 U_y (2 
A+\ell) \cos \theta +3 U_z (2 A+\ell) \sin \theta\right]\be_x.
\end{align}
Likewise, the hydrodynamic force exerted on rod $\#2$ is
\begin{dmath}
 \mathbf{F}^{\mathrm{hydro}}_{2} = \frac{2 \pi \mu \ell }{c}  \left[  
\theta_t  \lp A + \ell \rp \cos (\theta +2 \theta_1 )+3 \theta_t  (A+\ell) \cos 
\theta + 2 \ell 
(\theta_t +\theta_{1,t} ) \cos (\theta +\theta_1 )
+U_y \cos 2 (\theta +\theta_1 )+3 U_y+U_z \sin 2 (\theta 
+\theta_1 )\right]\be_y \\
+\frac{2 \pi \mu \ell}{c}  \left[  \theta_t \lp A + \ell \rp \sin (\theta +2 
\theta_1 ) + 3 
\theta_t  (A+\ell) \sin \theta+2 \ell  \lp\theta_t + \theta_{1,t}\rp \sin 
(\theta +\theta_1 
)+U_y \sin 2 (\theta +\theta_1 )-U_z\cos 2 (\theta +\theta_1 
)+3 U_z\right]\be_z
\end{dmath}
and the hydrodynamic torque on rod $\#2$ about $\Jone$ is
\begin{dmath}
 \boldsymbol{\Gamma}^{\mathrm{hydro}}_{2}|_{\Jone} = \frac{4 \pi   \mu 
\ell^2}{3 c} \left[3 \theta_t  (A+\ell) \cos \theta_1 +2 \ell \lp \theta_t + 
\theta_{1,t}\rp +3 U_y \cos (\theta +\theta_1 )+3 U_z \sin (\theta +\theta_1 
)\right].
\end{dmath}
The torque-free condition on rod $\#2$ reads
\begin{align}\label{eq:torque_rod2}
 \bM_{2} + \boldsymbol{\Gamma}^{\mathrm{hydro}}_{2}|_{\Jone} 
= \mathbf{0},
\end{align}
where $\bM_{2} = -K\theta_1 \be_x$ is the elastic moment exerted on rod $\#2$
by the torsional spring.
The torque balance on the whole composite system about the 
particle centre P is
\begin{align}\label{eq:torque_linker}
\boldsymbol{\Gamma}^{\mathrm{hydro}}_{1}|_{\mathrm{P}} + 
\underbrace{\lp \boldsymbol{\Gamma}^{\mathrm{hydro}}_{2}|_{\Jone} + 
\overrightarrow{\mathrm{P}\Jone}\times 
\mathbf{F}^{\mathrm{hydro}}_{2}\rp}_{\text{hydrodynamic torque 
on rod } \#2 \text{ about P}} - \gamma_{\mathrm{drag}} 
\theta_t\be_x + \underbrace{\lp 
E_3\mP_Q-E_Q\mP_3 
\rp \be_x}_{\text{electric torque on the particle}} = \mathbf{0}.
\end{align}
We also need to impose the force-free condition on the whole 
composite 
object
\begin{align}\label{eq:linker_force}
 \mathbf{F}^{\mathrm{hydro}}_{1} + \mathbf{F}^{\mathrm{hydro}}_{2} - 
\beta_{\mathrm{drag}} \lp U_y \be_y + U_z\be_z\rp = \mathbf{0}.
\end{align}
To close the system, we solve
the governing
equations~(\ref{eq:PQ}) and (\ref{eq:P3}) for $\mP_Q$ 
and $\mP_3$, where the second term $-\frac{\partial \psi}{\partial 
t}\mP_N$ in equation~(\ref{eq:PQ}) disappears.
We note that equations~(\ref{eq:torque_rod2}) and (\ref{eq:torque_linker}) 
indeed
reflect the subtle interplay between the elastic, electric and hydrodynamic 
torques, which lead to the EEH instability-induced self-oscillation.
\subsection{Nondimensionalization of the minimal model}
We use the same characteristic scales as the original particle-filament 
configuration (see \S~\ref{sec:setup_math}) to nondimensionalise 
equations~(\ref{eq:torque_rod2}), (\ref{eq:torque_linker}) and 
(\ref{eq:linker_force}), except that we substitute $D$
by $KL$, resulting in a slightly modified EEV parameter
\begin{align}
 \mbarmu = \frac{8\pi \mu L^3}{K\taus},
\end{align}
to be distinguished 
from $\barmu$ defined by equation~(\ref{eq:barmu}) for the original setup.
The dimensionless governing equations for  $\barU_y(\nont)$, 
$\barU_z(\nont)$, $\theta(\nont)$ and $\theta_{1}(\nont)$ are
\begin{subequations}\label{eq:non_model_vel}
 \begin{dmath}
  \lp 7\alpha+5/2 \rp \baromg\cos\theta + \baromg \lp 
\alpha+1/2 
\rp\cos\lp \theta+2\theta_1 \rp + \lp \baromg + \baromg_1 \rp 
\cos\lp \theta+\theta_1\rp + \barU_y\left[ \cos 
2\theta+\cos2\lp \theta+\theta_1 \rp - 6\alpha c +6 \right] + \barU_z \left[ 
\sin 2\theta+\sin2\lp \theta+\theta_1 \rp\right]   = 0,
 \end{dmath}
 \begin{dmath}
\lp 7\alpha+5/2 \rp \baromg \sin{\theta} + \baromg \lp \alpha+1/2 
\rp\sin\lp \theta+2\theta_1 \rp + \lp \baromg + \baromg_1 \rp 
\sin\lp \theta+\theta_1\rp
+\barU_z\left[ -\cos 
2\theta-\cos2\lp \theta+\theta_1 \rp - 6\alpha c +6  \right]
+ \barU_y \left[ \sin{2\theta} 
+ \sin2\lp \theta+\theta_1 \rp \right] 
=0,
 \end{dmath}
 \begin{dmath}
  \frac{\mbarmu}{24}\left[3\baromg \lp \alpha+1/2 \rp \cos\theta_1 + 
\baromg + \baromg_1 + 3\barU_y \cos \lp \theta+\theta_1 \rp 
 + 3\barU_z \sin\lp \theta+\theta_1 \rp \right] - c\theta_1 = 0, 
 \end{dmath}
 \begin{dmath}
  \frac{\mbarmu}{24}\left\{\lp21\alpha^2 +15\alpha+13/4 \rp\baromg 
+3\lp \alpha+1/2 \rp \lp \baromg + \baromg_1 \rp \cos\theta_1 \\
+3\lp 7\alpha+5/2 \rp \lp \barU_y\cos{\theta} + \barU_z \sin{\theta} \rp \\
+3\lp \alpha+1/2 \rp \cos2\theta_1 \left[ \lp \alpha+1/2 \rp \baromg + 
\barU_y\cos\theta +\barU_z \sin\theta \right]  + 3\lp \alpha+1/2 \rp 
\sin2\theta_1\lp \barU_z \cos\theta - \barU_y \sin\theta\rp  \right\} \\
 + c \theta_1 - c \bareta\baromg + c \barE \lp  \nP_Q\cos\theta  -  
  \nP_3 \sin \theta  
\rp = 0,
 \end{dmath}
\end{subequations}
where $c=1+2\log\esl$ and $\bareta = 
\alpha^3\mbarmu$ as given by
equations~(\ref{eq:c}) and (\ref{eq:bareta}), respectively.

The dimensionless equations for $\nP_Q$ and $\nP_3$ are
\begin{subequations}\label{eq:non_P_model}
 \begin{align}
 \frac{\partial \nP_Q}{\partial \nont} & = -\kappa \lp\nP_Q + \kappa \bareta 
\nE \sin\theta \rp,  \label{eq:non-PQ-model} \\
\frac{\partial \nP_3}{\partial \nont} & = -\kappa \lp \nP_3 + \kappa \bareta 
\nE \cos\theta\rp, \label{eq:non-P3-model}
 \end{align}
\end{subequations}
with their initial values at $\nont=0$
\begin{subequations}
\begin{align}
 \nP_Q \lp \nont = 0\rp & =  \frac{\bareta \kappa^2 \barE 
\sin\theta}{\kappa-\lp R-1 \rp/\lp 
S-1\rp}, \\
 \nP_3 \lp \nont = 0\rp & =  \frac{\bareta \kappa^2 \barE 
\cos\theta}{\kappa-\lp R-1 \rp/\lp 
S-1\rp}.
\end{align} 
\end{subequations}

\subsection{Numerical and theoretical (LSA) results of the minimal 
model}
\begin{figure}
\begin{center}
\hspace{0em}\includegraphics[width=1\textwidth]
{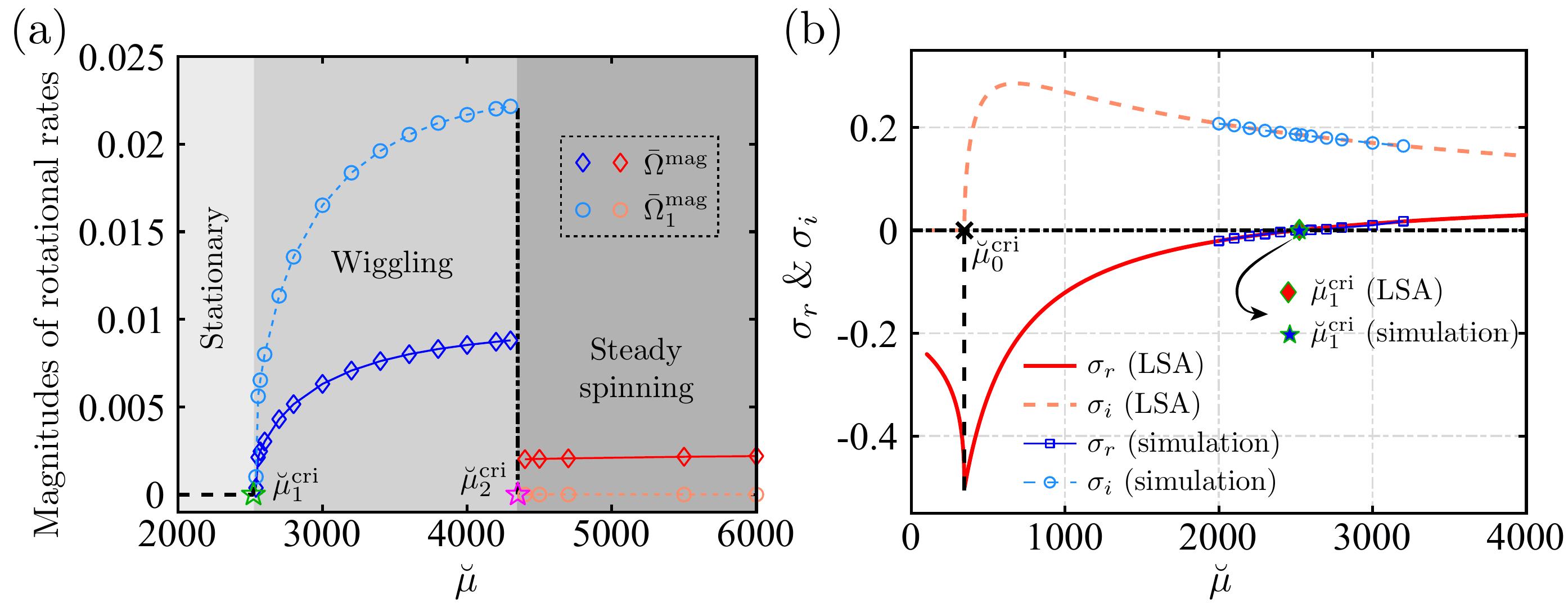}
\end{center}
\caption{(a) Numerical results of the rotational velocity magnitudes of the 
minimal model 
versus $\mbarmu$ by solving  equations (\ref{eq:non_model_vel}) and 
(\ref{eq:non_P_model}), where $\lp\barE,\alpha \rp=\lp 1.5, 
0.3\rp$; diamonds
and circles denote those of the particle and rod $\#2$, respectively.
$\mbarmucrione$ and $\mbarmucritwo$ separate the three $\mbarmu$-dependent 
regimes: stationary, wiggling (blue) and steady spinning (red). (b) 
LSA (red) and numerical (blue) results of the real $\sigr$ and imaginary 
$\sigi$ parts of the complex growth rate $\sigma$ versus $\mbarmu$, where 
$\lp\barE,\alpha \rp=\lp 1.5,0.3\rp$. $\mbarmucrizero$ distinguishes whether 
the 
perturbations decay monotonically when $\mbarmu < \mbarmucrizero$ or in an 
oscillatory manner when $\mbarmucrizero < \mbarmu <\mbarmucrione$.}
\label{fig:results-linker}
\end{figure}
We solve equations (\ref{eq:non_model_vel}) and 
(\ref{eq:non_P_model}) numerically using the MATLAB solver 
`ode15s' for ordinary differential equations. Fixing the electric field 
$\barE=1.5$
and size ratio $\alpha=0.3$, we show in 
figure~\ref{fig:results-linker}a
the $\mbarmu$-dependent magnitudes $\baromgamp$ and $\baromgamp_1$ 
 of the rotational velocities
of the particle and rod $\#2$, respectively, when the minimal composite object
reaches its equilibrium configuration. This simple model
reproduces the three
characteristic behaviours of the original particle-filament 
system: stationary ($\mbarmu<\mbarmucrione\approx2513$), wiggling 
($\mbarmucrione < \mbarmu < \mbarmucritwo \approx4300$)
and steady spinning ($\mbarmu>\mbarmucritwo$). 
In the spinning state, 
$\baromgamp_1 = |\mathrm{d} \theta_1 /\mathrm{d} \nont|=0$
reflects a time-independent angle $\theta_1$ between the two rods,
which adopt a steady ``deformed'' configuration
representing a minimal model of the deformed filament.

Conducting an LSA for this minimal model, we 
find the closed-form expression of the complex growth rate $\sigma=\sigr + 
\ii \sigi$ (see appendix~\ref{sec:lsa_mini} for details). The theoretical
values of $\sigma_{r,i}$ versus $\mbarmu$
in case of $\lp\barE,\alpha\rp=\lp1.5,0.3\rp$ are depicted in 
figure~\ref{fig:results-linker}b, as well as their numerical counterparts 
in the near $\mbarmucrione$ regime. The theoretical 
and numerical values of both $\sigr$ and $\sigi$ almost lie on 
top of each other, consequently, their predictions of 
$\mbarmucrione$ (when $\sigr=0$) agree. This superior 
agreement to the  
particle-filament system (figure~\ref{fig:sigma_v_mu}) is expected, because the 
minimal model does not 
require an approximate model (see \S \ref{sec:eh_model}) for the 
elastic torque as the original case.

The LSA also indicates the emergence of another subtle critical EEV number
$\mbarmucrizero \approx 345$ (black cross in figure~\ref{fig:results-linker}b): 
when $\mbarmu<\mbarmucrizero$, the real part $\sigr$ of the growth rate is 
negative, accompanying a zero imaginary part, thus the 
perturbations diminish  to zero monotonically; when $\mbarmucrizero < 
\mbarmu < \mbarmucrione$, $\sigr<0$ but $\sigi>0$, the perturbations also 
die out but in a an oscillatory fashion. The former case corresponds to the 
non-negative quantity
$\Sigma$ inside the square-root operator in equation~(\ref{eq:sigma_linker}) 
that naturally yields real solutions for $\sigma$ only. A similar structure of 
the 
solutions of $\sigma$ was reported in \citet{de2017spontaneous}.
Since the current work mainly addresses the EEH instability-induced 
self-oscillation, 
we do not 
pursue a detailed investigation in this stable, stationary regime.

\section{Conclusions and discussions}\label{sec:conclusions}
Standard biomimetic practises commonly rely on an oscillating magnetic 
or electric field to produce the oscillatory motion of slender artificial 
structures. In contrast, we propose a strategy to achieve 
self-oscillation of artificial structures based on a time-independent,
uniform electric field. By formulating and numerically solving an 
elasto-electro-hydrodynamic problem, this concept is illustrated
by oscillating a composite object consisting of a weakly conducting dielectric 
spherical particle and an elastic filament immersed in a dielectric solvent.

Our strategy is grounded in the QR electrohydrodynamic instability 
phenomenon indicating that a weakly conducting dielectric particle suspended 
in a dielectric liquid of higher conductivity can undergo spontaneous rotation
under a sufficiently strong DC electric field. For an individual spherical 
particle, this instability emerges through a supercritical pitchfork 
bifurcation resulting in steady rotation~\citep{jones1984quincke}. 
By incorporating an elastic filament, we transform the 
pitchfork bifurcation into a Hopf 
bifurcation through which a self-oscillatory 
instability occurs~\citep{zhu2019propulsion}. 
This transformation is attributed to the elasto-viscous 
response of the filament providing an elastic torque to balance
the electric and hydrodynamic torques. The elastic torque is in phase with 
the 
rotational velocity of the particle at certain time periods (see 
figure~\ref{fig:torque_omg}b). This in-phase 
behaviour results in negative
damping (or positive feedback), hence leading to the onset of linear 
instability~\citep{jenkins2013self}. We comment that such a transition from 
pitchfork to Hopf bifurcation was also identified 
by \citet{tsebers1980electrohydrodynamic} who observed  
oscillatory QR of ellipsoidal particles attributed to their anisotropic 
electric properties.
It is also worth mentioning that the QR
instability was utilised to study suspensions of artificial swimmers made of
QR particles that achieved locomotion by rolling near a rigid solid
boundary~\citep{bricard2013emergence}. In addition, the  recent work of 
\citet{das2019active} shows theoretically and 
numerically that a dielectric particle with particular geometrical asymmetry 
(e.g. a helix) under a DC electric field is able to convert QR into spontaneous 
translation in an unbounded domain.

We next recall the original experiments conducted
by~\citet{quincke1896ueber}, where the particle was hung by a silk 
thread and hence the particle rotated in the direction along the orientation of 
the thread. Quincke also noted an oscillatory behaviour as translated 
by~\citet{jones1984quincke}\\

``\textit{Quincke, with his spheres
tethered to silk threads, had been forced to contend with
periodic rotation, first in one direction and then in the other as
the silk thread wound and unwound}''.\\

We think that the ``wound and unwound'' motion  
manifested the self-oscillatory phenomenon, which is attributed 
to the torsional deformation of the silk thread. 
We speculate that Quincke probably regarded this observation as an 
experimental nuisance, thus did not pay attention to it nor did other 
researchers, 
except for one little-known preprint~\citep{zaks_onset} that recognised 
and modelled this torsional oscillation by considering
a QR particle hung by a thread with torsional elasticity.

In this paper, we consider only the bending stiffness of the grafted
filament and the whole composite object is freely suspended in the solvent.
By applying an electric field stronger than 
the critical value corresponding to the onset of original QR instability, 
the composite object exhibits three distinct behaviours depending on the EEV 
number $\barmu$ (inversely proportional to the bending stiffness). When $\barmu 
\leq \barmucrione$, the object remains stationary, corresponding to a 
fixed-point solution; when $\barmu \geq \barmucritwo$, the particle spins 
steadily towing a deformed filament, corresponding to an 
asymmetric fixed-point solution; when $\barmu \in \lp \barmucrione, 
\barmucritwo 
\rp$, the particle oscillates
and the filament wiggles, leading the object to an undulatory locomotion.
More specifically, instability occurs at $\barmucrione$ through a 
supercritical Hopf bifurcation, where the self-oscillatory motion 
represents a limit-cycle solution; at $\barmucritwo$, a secondary
bifurcation appears, and the oscillatory, limit-cycle solution jumps to 
the steadily spinning, fixed-point solution. By fixing the EEV number $\barmu$,
bifurcation diagrams considering the electric field strength $\barE$
as the control parameter revealed the same three scenarios (see 
figure~\ref{fig:omg_v_E_varyMu_a0.3}).

We have also examined the propulsive performance of the micro object
in the self-oscillating regime $\barmu \in (\barmucrione, \barmucritwo)$.
The trajectory of the object resembles a wave propagating along a straight
path. The translational velocity of the object along this path
varies in $\barmu$ non-monotonically (see figure~\ref{fig:opt_mu_and_spd}c).

Motivated by the exponential temporal growth of the rotational velocity, 
we 
performed a LSA to predict theoretically the onset of the self-oscillatory 
instability.
We have developed an elastohydrodynamic model to account for the elastic 
force and torque exerted by the filament on the particle, which  
closely matched the numerical counterparts. Incorporating this 
model into a standard LSA for the original QR particle, we derived 
the dispersion relationship of the new EEH problem. We thus calculated the 
complex growth rate $\sigma=\sigma_r + \ii\sigma_i$ and identified the critical 
EEV number $\barmucrione$. Theoretical predictions of $\sigma$ 
(figure~\ref{fig:sigma_v_mu}) and $\barmucrione$ (figure~\ref{fig:crimu_v_E})
agree well the numerical results, especially in the large $\barmu$ regime. 
However, the agreement becomes less 
satisfactory when $\barmu$ decreases because of the violation of  an 
assumption used in the elastohydrodynamic model.

To unravel the EEH instability mechanism, we studied a minimal model system
characterised by two rigid rods linked by a torsional spring to mimic the 
original filament. This substitution reduces the elastic element's number of 
degrees of freedom to one. Numerical and LSA results demonstrated that the 
minimal model could exhibit the three elasticity-dependent behaviours: 
stationary, wiggling and steady spinning.

Following the comments of an anonymous referee, we hereby emphasise the 
 difference between our work and other seemingly similar 
studies~\citep{manghi2006propulsion,qian2008shape,coq}, where a
flexible slender structure (filament or rod) rotated in a viscous fluid and 
produced thrust because 
one of its ends was clamped to a constantly rotating base or actuated by a 
constant torque.
This rotation results from 
forced oscillation characterised
by a close correlation between the frequency of the power source and that of 
the resulting periodic motion. This forced-oscillatory periodic motion 
distinguishes itself from the self-oscillatory motion we observe, where the 
time-independent electric field as the power source lacks a frequency
corresponding to that of the periodic motion.

The current work constrained the kinematics and 
electric polarisation vector of the particle to a plane in order to show a 
clean physical picture of the new EEH instability we identified. By removing 
these constraints, we anticipate the appearance of more complex and diverse  
three-dimensional 
behaviours featured by bi/multi-stability, hysteresis and even chaos 
(ellipsoidal particles were observed to exhibit chaotic 
QR~\citep{tsebers1991chaotic}). We will report the results of the ongoing work 
in a future paper.

It is also worth mentioning the assumption of neglecting 
electrohydrodynamic 
effect of the filament. The electric torque exerted on a slender QR structure 
scales with $a^2 L$~\citep{das2019active}, and that on a sphere scales with 
$A^3$ (see equation~(\ref{eq:torque_qr_sphere})). By assuming that the filament
and particle have similar dielectric properties and realising $\alpha=A/L=
O(1)$, the ratio of the former to the latter torque
is of the order of $\esl^2$. This comparison thus justifies the assumption, 
which 
also implies that no special attention needs to be paid in this context for the 
experimental realisation.

In conclusion, incorporating an elastic element to manipulate the 
electrohydrodynamic 
instability, we report an elasto-electro-hydrodynamic instability and use it
for engineering self-oscillation of artificial structures. 
We anticipate that this idea of harnessing elastic media to control and 
diversify the bifurcation and the corresponding 
instability behaviour can be generalised to
other stability phenomena and systems.
As a result, different emerging instability behaviours can 
be utilised for diverse functionalities. This concept might inspire new
approaches to design soft, reconfigurable machines that can morph and 
adapt to the environment.

Declaration of Interests. The authors report no conflict of interest.

\appendix

\section{Numerical methods for the EEH problem}\label{sec:numerical}
In this section, we describe the numerical methods to solve the EEH problem.
For notation brevity, we remove all the bars over the unknown dimensionless
variables henceforth. Following \cite{tornberg2004simulating}, we use a 
finite difference 
scheme to discretise the filament centreline by a uniform grid of 
$N$ points. A typical value of $N=201$ is used in the simulations.
Therefore, we have $N$ unknowns for the tension 
$T(s)$ and $3N$ for the coordinates $\br(s)=x \be_x + y \be_y + z 
\be_z$. The implementation considers a general three-dimensional motion of the 
filament, hence we also solve for $x (s)$ even though the motion  is 
restricted to
the $yz$-plane.  

In contrast, for the restricted motion of the particle, 
we express its translational velocity $\bU$ and rotational velocity 
$\bOmega$ as
\begin{subequations}
\begin{align}
 \bU &=  U_y\be_y + U_z\be_z, \\
 \bOmega & = \Omega \be_x = \theta_t \be_x,
\end{align} 
\end{subequations}
which yields three unknowns
$U_y$, $U_z$ and $\Omega=\theta_t$ for the particle. 
Another two unknowns are $\mP_Q$ and $\mP_3$ governed by
equations (\ref{eq:non-PQ}) and (\ref{eq:non-P3}), respectively. 

At the $(k+1)$-th time step, $T^{k+1}(s)$ is solved based on 
$\br^{k}(s)$ at the 
$k$-th time step. We then solve $[\br^{k+1}(s), U^{k+1}_y, U^{k+1}_z, 
\Omega^{k+1}, \mP^{k+1}_Q, \mP^{k+1}_3]$ in a coupled way, which consists of 
$3N+5$ unknowns. We adopt this coupled strategy to accurately preserve  
the clamped BC of the filament base J ($s=0$): first, the filament base
is on the particle surface; second, the tangent vector $\br_s|_{s=0}$ at the 
base always
passes through the particle centre P. This clamped BC of the filament  is 
different from other 
configurations~\citep{guglielmini2012buckling,de2017spontaneous} where the base 
is stationary. In our case, the filament base is 
right on the particle surface, and is able to translate with and rotate about 
the particle centre. 

We use the backward Euler formulation to approximate the particle centre 
$\bxp^{k+1}$, 
\begin{align}
 \bxp^{k+1} = \bxp^{k} + \Delta t  \bU^{k+1},
\end{align}
where $\Delta t$ is the time step. The Dirichlet BC for $\br|_{s=0}$, namely 
equation~(\ref{eq:nonbr_s0}) becomes
\begin{align}\label{eq:num_bc_s0_1}
 \br^{k+1} & = \bxp^k + \Delta t \bU^{k+1} + \alpha \br_s^{k+1}. 
\end{align}
Likewise, we write  $\theta^{k+1} = \theta^k + \Delta t \Omega^{k+1}$.
By assuming $\Delta t \Omega \ll 1$, we obtain
\begin{subequations}\label{eq:theta_small}
 \begin{align}
 \cos{\theta^{k+1}} & \approx \cos{\theta^{k}}- \lp \dt \sin{\theta^{k}} \rp 
\Omega^{k+1}, \\
 \sin{\theta^{k+1}} & \approx \sin{\theta^{k}}+ \lp \dt \cos{\theta^{k}} \rp 
\Omega^{k+1}. 
\end{align}
\end{subequations}
The tangent vector $\br^{k+1}_s$ at the filament base is opposite to 
the particle orientation $\ortp$, namely $\br^{k+1}_s = 
\sin{\theta^{k+1}}\be_y - \cos{\theta^{k+1}}\be_z$. Using
equation~(\ref{eq:theta_small}), the discretised form of this BC is,
\begin{align}\label{eq:num_fix_tangent}
 \br_s^{k+1} - \lp \dt \cos{\theta^k} \rp \Omega ^{k+1} \be_y - \lp 
\dt \sin{\theta^k} \rp \Omega ^{k+1} \be_z &= \sin{\theta^k}\be_y - 
\cos{\theta^k}\be_z.
\end{align}

Combining equation~(\ref{eq:force-free-non}) and 
(\ref{eq:non_elastic_force}),
the discretised form for the force-free condition reads
\begin{subequations}\label{eq:num_force_free}
 \begin{align}
  -y^{k+1}_{sss}|_{s=0} + \lp T^{k+1}y^{k+1}_{s} \rp|_{s=0} - 
3\alpha\barmu U_y^{k+1}/ 4  & =0,\\
  -z^{k+1}_{sss}|_{s=0} + \lp T^{k+1}z^{k+1}_{s} \rp|_{s=0} - 
3\alpha\barmu U_z^{k+1}/4  & =0.
 \end{align}
\end{subequations}
Combining equation~(\ref{eq:torque-free-non}) and (\ref{eq:non_elastic_torque}) 
similarly,
the torque-free condition  reads
\begin{align}\label{eq:num_torque_free}
 \bareta \Omega = \barE \cos{\theta} \mP_Q - \barE 
\sin{\theta} \mP_3 + \left[ \lp \sin\theta \be_y - \cos\theta \be_z \rp \times 
\lp \br_{ss} - \alpha \br_{sss} \rp \right]|_{s=0}.
\end{align}
We approximate $\theta^{k+1}$ by $\check{\theta} = 2\theta^{k}-\theta^{k-1}$
and substitute it into equation~(\ref{eq:num_torque_free}), deriving the 
discretised torque-free condition 
\begin{align}\label{eq:num_torque_free_dis}
 \bareta \Omega^{k+1} = \barE \cos{\check{\theta}} \mP^{k+1}_Q - \barE 
\sin{\check{\theta}} \mP^{k+1}_3 + \left[ \lp \sin\check{\theta} \be_y - 
\cos\check{\theta} \be_z \rp \times 
\lp \br^{k+1}_{ss} - \alpha \br^{k+1}_{sss} \rp \right]|_{s=0}. 
\end{align}

Using the backward Euler scheme for $\mP_Q$ and $\mP_3$, and combining
equations~(\ref{eq:theta_small}), we find the discretised 
 governing equations for $\mP_Q$ and $\mP_3$,
\begin{subequations}\label{eq:num_PQ3}
 \begin{align}
  \lp \kappa+1/\dt \rp \mP_Q^{k+1} + \kappa^2 \bareta \barE \dt 
\cos\theta^k \Omega^{k+1} & = \mP_Q^{k}/\dt - \kappa^2 \bareta \barE 
\sin\theta^k,\\
  \lp \kappa+1/\dt \rp \mP_3^{k+1} - \kappa^2 \bareta \barE \dt 
\sin\theta^k \Omega^{k+1} & = \mP_3^{k}/\dt - \kappa^2 \bareta \barE 
\cos\theta^k.
 \end{align}
\end{subequations}

Integrating equations (\ref{eq:fila_pos}), (\ref{eq:num_force_free}), 
(\ref{eq:num_torque_free}) and (\ref{eq:num_PQ3}), with the four BCs 
equations~(\ref{eq:BC_s1_r}), (\ref{eq:num_bc_s0_1}) and 
(\ref{eq:num_fix_tangent})
for $\br^{k+1}$ generates a linear system of size $3N+5$. Its solution 
corresponds to
$[\br^{k+1}(s), U^{k+1}_y, U^{k+1}_z, 
\Omega^{k+1}, \mP^{k+1}_Q, \mP^{k+1}_3]$.

\section{Quincke rotation of a dielectric sphere jointed with a rigid 
rod}\label{sec:append}
Following \citet{jones1984quincke}, we derive the critical electric 
field required to trigger the electrohydrodynamic instability of a dielectric 
spherical particle grafted by a rigid rod. Let us first briefly reproduce the 
derivation of \citet{jones1984quincke} for an individual particle and then 
extend it to our composite particle-rod configuration.

The electric torque exerted on a spherical particle of radius $A$ about its 
centre P  is
\begin{align}\label{eq:torque_qr_sphere}
 \bT^{\mathrm{elec}} & = \frac{6 \pi \es A^3 E^2 \lp 1-R/S \rp \taumw \Omega 
}{\lp 
1+\frac{2}{S} \rp \lp 1+\frac{R}{2} \rp \left[ 1+ \Omega^2 {(\taumw)}^2 
\right]} \be_x.
\end{align}
When the particle rotates about its centre at velocity $\Omega \be_x$,
the hydrodynamic torque exerted on it is
\begin{align}
 \bT_{\mathrm{par}}^{\mathrm{hydro}} & = -8\pi \mu A^3 \Omega \be_x.
\end{align}
By using the torque-free condition $\bT^{\mathrm{elec}} + 
\bT^{\mathrm{hydro}}_{\mathrm{par}}  = \mathbf{0}$, we derive
\begin{align}\label{eq:omega_tau}
  \Omega^2 {(\taumw)}^2 & = \frac{ 3 \es E^2 \lp 1 - R/S \rp \taumw}{4 \mu \lp 
1+\frac{2}{S} \rp \lp 1+\frac{R}{2} \rp}-1.
\end{align}
Because the left-hand side of equation~(\ref{eq:omega_tau}) is non-negative for
a real value of $\Omega$, this condition gives us the criterion of 
the electrical field $E$ above which QR instability occurs, 
\begin{align}\label{eq:app_ecri}
 E \geq \Ecri = \sqrt{\frac{2 \ss \mu (R+2)^2}{3 \es^2(S-R)}}.
\end{align}
The rotational speed
of the QR particle is known based on equation~(\ref{eq:omega_tau}), so that its
dimensionless value is
\begin{align}\label{eq:non_omega_QR}
  \nonOmega_{\mathrm{QR}} =  \kappa \sqrt{\barE^2-1},
\end{align}
where $\kappa=(R+2)/(S+2)$ as defined in equation~(\ref{eq:kappa}).

\begin{figure}
\begin{center}
\hspace{0em}\includegraphics[width=0.5\textwidth]
{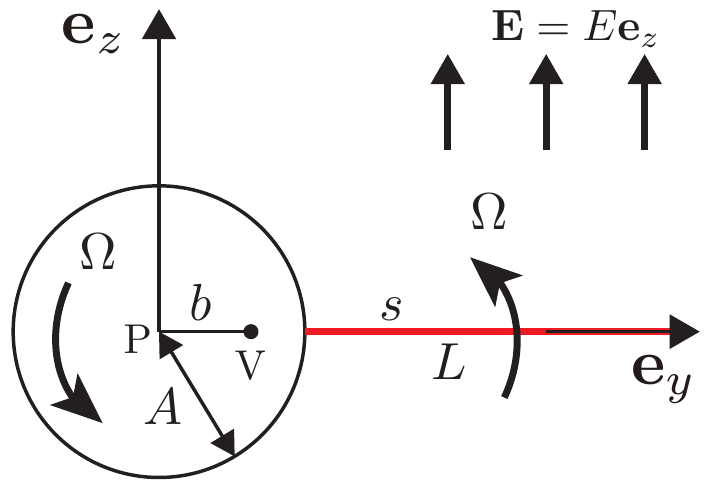}
\end{center}
\caption{Schematic showing a dielectric particle of radius 
$A$ connected by a rigid rod of length $L$ and radius $a$. Under a sufficient 
strong electric field $\bE = E\be_z$, the composite object rotates about a 
pivot 
V that lies on
the centreline of rod.}
\label{fig:sketch-rigid-rod}
\end{figure}

Now we adapt the above derivation to the composite particle-rod system steadily 
rotating at velocity $\bOmega = \Omega \be_x$ about an off-centre pivot point V 
on the 
$y$-axis (see figure~\ref{fig:sketch-rigid-rod}). We choose the particle centre 
P as the origin of the Cartesian coordinates.
Using the local SBT, the force per unit length $-\bf_{\mathrm{rod}}$ exerted by 
the 
fluid onto the rod at arclength $s$ is
\begin{align}
 -\bf_{\mathrm{rod}}(s) = -\xi_{\perp} \Omega \lp 
A- b +s \rp \be_z,
\end{align}
where $\xi_{\perp} = -8\pi\mu/c = -8\pi\mu/(1+2\log{\esl})$,
and the total hydrodynamic force $\bF_{\mathrm{rod}}$ exerted on the rod is
\begin{align}
 \bF_{\mathrm{rod}} = \int_{0}^{L} -\bf_{\mathrm{rod}}(s) \d s = -\xi_{\perp} 
\Omega   \left[ (A-b)L + L^2/2 \right] \be_z.
\end{align}
Simultaneously, the hydrodynamic force exerted on the translating 
spherical particle is 
$\bF_{\mathrm{par}} = 6\pi \mu A b \Omega \be_z$. Using the force-free 
condition $\bF_{\mathrm{rod}} + \bF_{\mathrm{par}} = \mathbf{0}$, we obtain
\begin{align}\label{eq:b_rod}
 b = \frac{\xi_\perp L \lp 1+2\alpha \rp}{2\lp 6\pi \mu \alpha + \xi_{\perp} 
\rp}.
\end{align}

The hydrodynamic torque $\bT^{\mathrm{hydro}}_{\mathrm{rod}}$ exerted on 
the 
rod 
with respect to the particle centre P is 
\begin{align}
 \bT^{\mathrm{hydro}}_{\mathrm{rod}} & = \int_{0}^{L} (A+s)\be_y \times 
\left[ 
-\xi_{\perp} \Omega 
\lp  A -b +s\rp \be_z \right]\d s \nonumber \\
& = -\xi_{\perp}\Omega \be_x \int_{0}^{L} \lp A+s \rp \lp A-b+s \rp \d s 
\nonumber \\
& = -\frac{\xi_{\perp}\Omega \be_z}{6} \left[ 6\lp A^2-Ab \rp L + 3\lp 2A-b \rp 
L^2 + 2L^3 \right].
\end{align}
Using the torque-free condition on the particle-rod system, 
$\bT^{\mathrm{elec}} 
+ \bT^{\mathrm{hydro}}_{\mathrm{par}} +  
\bT^{\mathrm{hydro}}_{\mathrm{rod}}   
= 
\mathbf{0}$, we obtain
\begin{align}
  \frac{\bT^{\mathrm{elec}} 
+ \bT^{\mathrm{hydro}}_{\mathrm{par}} +  \bT^{\mathrm{hydro}}_{\mathrm{rod}}}{ 
2 
\pi 
A^3 \Omega \be_x } = 
\frac{3 \es E^2 \lp 1-R/S 
\rp \taumw }{\lp 
1+\frac{2}{S} \rp \lp 1+\frac{R}{2} \rp \lp 1+ \Omega^2 {(\taumw)}^2 \rp} - 
4 \mu \left[ 1 + F\lp \alpha, \hat{\beta}, \esl \rp \right],
\end{align}
where 
\begin{subequations}
\begin{align}
 F\lp \alpha,\hat{\beta},\esl \rp & = \frac{
6\alpha^{-1} \lp 1 - \hat{\beta}^{-1} \rp 
+ 3\alpha^{-2} \lp 2-\hat{\beta}^{-1} \rp  + 2\alpha^{-3} }{6 \lp -1 - 
2\log{\epsilon} \rp}, \\
 \hat{\beta} =  A/b & = \frac{2\alpha}{4\lp 2\alpha+1 \rp} \left[ 4 - 3 
\alpha 
\lp  1 + 
2\log{\esl}\rp \right].
\end{align}
\end{subequations}
Hence, the critical electrical field corresponding to the instability inception
is
\begin{align}
 \bar{\mathcal{E}}^{\mathrm{cri}}  = \mathcal{E}^{\mathrm{cri}}/\Ecri = 
\sqrt{1+F}.
\end{align}
The typical values
of $\bar{\mathcal{E}}^{\mathrm{cri}}$ as a function of size ratio $\alpha=A/L$ 
for 
$\esl=0.01$ are provided in
table~\ref{tab:Ecri_rod}.

\begin{table}
\centering
\begin{tabular}{|c|c|c|c|c|c|}
\hline
 $\alpha$ & 0.1 &  0.3 & 0.5 & 0.7 & 0.9  \\ \hline
$\bar{\mathcal{E}}^{\mathrm{cri}}$ & 5.278 & 1.803 & 1.348 & 1.202 & 1.136 \\ 
\hline
\end{tabular}
\caption{Dimensionless critical electric field 
$\bar{\mathcal{E}}^{\mathrm{cri}}$ above which
the composite object of a dielectric sphere of size ratio $\alpha$ and a rigid 
rod undergoes the electrohydrodynamic instability, the slenderness 
$\esl=0.01$.}
\label{tab:Ecri_rod}
\end{table}

\section{Two-dimensional polynomial equations for $\hsig_r$ and 
$W$}\label{sec:polynomial}
We substitute equations~(\ref{eq:beta}) and (\ref{eq:log_xi}) into 
equation~(\ref{eq:hsig}). We define $\Ep = E^2-1$ and $\upsilon = \lp 
\frac{1}{-1-2\log{\esl}}\rp^{1/4}$ that leads to $\log\xi = z_0 
\upsilon W $. Consequently, the system of
two-dimensional polynomial equations for $\hsig_r$ and $W$ reads,
\begin{align}
\begin{split}
& 0  =  -9 \hsigr^2 W^8 \alpha^5-24\hsig^2 W^7 
\alpha^3\upsilon^3\lp\cos\frac{3\pi}{8}+\sin\frac{3\pi}{8} \rp -32\hsigr^2W^6
\alpha^3\upsilon^6\sin\frac{3\pi}{4} \\
& -57\hsigr W^{11} \alpha^4\upsilon^3\lp -\cos\frac{3\pi}{8} 
+\sin\frac{3\pi}{8}  \rp -2\hsig W^{10}\alpha^3 \upsilon^2 \lp -44\upsilon^4 
\cos{\frac{3\pi}{4}} + 9\sin\frac{\pi}{4}\rp\\
&-3\hsigr W^9 \alpha^2\upsilon \left[3\cos{\frac{\pi}{8}} +3\sin{\frac{\pi}{8}} 
+8\upsilon^4 \lp -\cos{\frac{5\pi}{8}} +\sin{\frac{5\pi}{8}} \rp \right] + 
9\hsigr W^8\alpha \lp -4\upsilon^4 +\kappa \barmu\alpha^4 \Ep\rp\\
&+8\hsigr W^7 \upsilon^3 \left[ 2\upsilon^4\lp \cos{\frac{7\pi}{8}} 
-\sin{\frac{7\pi}{8}}  \rp + 3\kappa \barmu\alpha^4\Ep\lp \cos{\frac{3\pi}{8}} 
+\sin{\frac{3\pi}{8}} \rp  \right] + 32\kappa \barmu\hsigr W^6 \alpha^3 
\upsilon^6 \sin{\frac{3\pi}{4}}\Ep \\
&+9 W^{16} \alpha^5 + 33W^{15}\alpha^4\upsilon^3\lp\cos{\frac{3\pi}{8}} 
+\sin{\frac{3\pi}{8}} \rp + 2W^{14}\alpha^3\upsilon^2\lp 9\cos{\frac{\pi}{4}} 
+28\upsilon^4 \sin{\frac{3\pi}{4}} \rp\\
&+3W^{13}\alpha^2\upsilon \left[ 3\cos{\frac{\pi}{8}} - 3\sin{\frac{\pi}{8}} + 
8\upsilon^4 \lp \cos{\frac{5\pi}{8}} +\sin{\frac{5\pi}{8}} \rp \right] \\
&+W^{11}\upsilon^3\left[ 9\kappa \barmu\alpha^4\lp \cos{\frac{3\pi}{8}} 
-\sin{\frac{3\pi}{8}} \rp  + 16\upsilon^4 \lp \cos{\frac{7\pi}{8}} 
+\sin{\frac{7\pi}{8}} \rp - 24\kappa \barmu\alpha^4\Ep \lp \cos{\frac{3\pi}{8}} 
-\sin{\frac{3\pi}{8}} \rp \right] \\
&-2\kappa \barmu W^{10} \alpha^3 \upsilon^2 \lp 
-12\upsilon^4\cos{\frac{3\pi}{4}} +9\sin{\frac{\pi}{4}} + 16\upsilon^4 \Ep 
\cos{\frac{3\pi}{4}}\rp \\
&+ 
3\kappa \barmu W^9\alpha^2\upsilon\left[-3\lp \cos{\frac{\pi}{8}} 
+\sin{\frac{\pi}{8}} \rp  + 8\upsilon^4 \lp \cos{\frac{5\pi}{8}} 
-\sin{\frac{5\pi}{8}} \rp \right] \\
&-36 \kappa \barmu W^8 \alpha\upsilon^4 +16 \kappa \barmu W^7 \upsilon^7 \lp 
\cos{\frac{7\pi}{8}} -\sin{\frac{7\pi}{8}} \rp,
\end{split}
\end{align}
and
\begin{align}
 \begin{split}
& 0  =  -24\hsig^2W^7\alpha^4\upsilon^3\lp \cos{\frac{3\pi}{8}} 
-\sin{\frac{3\pi}{8}} \rp - 32\hsig^2 W^6 \alpha^3 \upsilon^6 
\cos{\frac{3\pi}{4}} - 18\hsig W^{12}\alpha^5\\
&-57\hsig W^{11}\alpha^4 \upsilon^3 \lp \cos{\frac{3\pi}{8}} 
+\sin{\frac{3\pi}{8}} \rp - 2\hsig W^{10}\alpha^3\upsilon^2 \lp 
9\cos\frac{\pi}{4} + 44\upsilon^4 \sin{\frac{3\pi}{4}} \rp\\
&-3\hsig W^9 \alpha^2 \upsilon \left[ 3\lp \cos{\frac{\pi}{8}} 
-\sin{\frac{\pi}{8}} \rp  + 8\upsilon^4 \lp \cos{\frac{5\pi}{8}} 
+\sin{\frac{5\pi}{8}} \rp \right]\\
&-8\hsigr W^7\upsilon^3 \left[ 2\upsilon^4 \lp\cos{\frac{7\pi}{8}} 
+\sin{\frac{7\pi}{8}}\rp - 3\kappa \barmu\alpha^4 \Ep \lp \cos{\frac{3\pi}{8}} 
-\sin{\frac{3\pi}{8}}  \rp \right] + 32 \kappa \barmu \hsigr W^6 \alpha^3 
\upsilon^6 
\Ep \cos{\frac{3\pi}{4}} \\
&+33W^{15}\alpha^4\upsilon^3\lp \cos{\frac{3\pi}{8}} -\sin{\frac{3\pi}{8}} 
\rp+2W^{14}\alpha^3\upsilon^2\lp 28\upsilon^4
\cos{\frac{3\pi}{4}} - 9\sin\frac{\pi}{4}\rp \\
&+3W^{13}\alpha^2\upsilon \left[-3\lp \cos{\frac{\pi}{8}} 
+\sin{\frac{\pi}{8}} \rp  + 8\upsilon^4 \lp \cos{\frac{5\pi}{8}} 
-\sin{\frac{5\pi}{8}} \rp \right]+9W^{12}\alpha\lp  
-4\upsilon^4+\kappa \barmu\alpha^4\Ep\rp \\
&+W^{11}\upsilon^3\left[-9\kappa \barmu\alpha^4\lp \cos{\frac{3\pi}{8}} 
+\sin{\frac{3\pi}{8}} \rp  + 16\upsilon^4 \lp \cos{\frac{7\pi}{8}} 
-\sin{\frac{7\pi}{8}} \rp +24\kappa \barmu\alpha^4\Ep \lp \cos{\frac{3\pi}{8}} 
+\sin{\frac{3\pi}{8}} \rp \right]\\
&+2\kappa \barmu W^{10}\alpha^3\upsilon^2\lp -9\cos{\frac{\pi}{4}} 
-12\upsilon^4 \sin\frac{3\pi}{4} + 16\upsilon^4\Ep \sin{\frac{3\pi}{4}}\rp\\
&-3\kappa \barmu W^9 
\alpha^2\upsilon \left[ 3\lp \cos{\frac{\pi}{8}} -\sin{\frac{\pi}{8}} \rp 
 + 8\upsilon^4 \lp \cos{\frac{5\pi}{8}} +\sin{\frac{5\pi}{8}} \rp \right] 
-16\kappa \barmu W^7 \upsilon^7 \lp \cos{\frac{7\pi}{8}} +\sin{\frac{7\pi}{8}} 
\rp.
 \end{split}
\end{align}

\section{LSA for the minimal model}\label{sec:lsa_mini}
We perform LSA for the minimal model.  Similar to 
\S~\ref{sec:linear_qc}, the bars over the dimensionless unknown variables are 
dropped, unless otherwise 
specified. The state variables $[\theta,\theta_1, U_y, U_z, \mP_Q, \mP_3]$
are decomposed into a  base state $[\bstheta,\bstheta_1, 
\hat{U}_y, \hat{U}_z, \hat{\mP}_Q, \hat{\mP}_3]$ and a perturbation state 
 $[\ptheta, \pthetaone, \pUy, \pUz, \pPQ, \pPT]$, where $\bstheta_1 = \hat{U}_y 
= 
\hat{U}_z = 0$, $\bstheta$ is an arbitrary value, and $\hat{\mP}_Q$ and 
$\hat{\mP}_3$  are given by equation~(\ref{eq:base-dipole}).
By linearising equations (\ref{eq:non_model_vel}) and 
(\ref{eq:non_P_model}) with respect to the base state, we derive the linear 
evolution
equations for the perturbative state variables,
 \begin{align}\label{eq:mini_gov_pert}
 (8\alpha+4) \frac{\mathrm{d} \ptheta}{\mathrm{d} t} \cos\bstheta  
+\frac{\mathrm{d} \pthetaone}{\mathrm{d} t}  \cos\bstheta + 2\lp  3 - 3\alpha c 
+ \cos 2\bstheta \rp \pUy + 2 \pUz \sin2\bstheta  & = 0, \nonumber \\
 (8\alpha+4) \frac{\mathrm{d} \ptheta}{\mathrm{d} t} \sin\bstheta  
+\frac{\mathrm{d} \pthetaone}{\mathrm{d} t} \sin\bstheta  + 2\lp  
 3 - 3\alpha c -\cos2\bstheta \rp \pUz + 2 \pUy \sin2\bstheta  & = 0,\nonumber 
\\
 \lp 3\alpha+5/2 \rp\barmu \frac{\mathrm{d} \ptheta}{\mathrm{d} t} + \barmu 
\frac{\mathrm{d} \pthetaone}{\mathrm{d} t}+ 3\barmu \pUy \cos\bstheta+3\barmu 
\pUz \sin\bstheta - 24c \pthetaone & =0,\nonumber \\
 \lp 24\alpha^2+21\alpha+11/2 \rp \barmu \frac{\mathrm{d} \ptheta}{\mathrm{d} 
t} 
+ 3\lp  \alpha + 1/2\rp \barmu \frac{\mathrm{d} \pthetaone}{\mathrm{d} t} + \lp 
24\alpha + 9 \rp \barmu \lp \pUy \cos\bstheta + \pUz \sin \bstheta \rp  
\nonumber \\
+ 24 c \pthetaone - 24 c \bareta \frac{\mathrm{d} \ptheta}{\mathrm{d} t} + 24 c 
\bareta \kappa \barE^2  \ptheta 
+ 24 c\barE \pPQ 
\cos\bstheta - 24 c \barE \pPT\sin\bstheta & =0 \nonumber \\
\frac{\partial \pPQ}{\partial t}  +\kappa \pPQ + \kappa^2 \bareta \barE 
\ptheta \cos\bstheta &= 0,\nonumber \\
\frac{\partial \pPT}{\partial t} + \kappa \pPT - \kappa^2 \bareta \barE 
\ptheta \sin\bstheta &=0.
 \end{align}
Employing the normal-mode approach as in \S{~\ref{sec:linear}}, we  
substitute $\left[\ptheta, \pthetaone, \pUy, \pUz, \pPQ, \pPT \right] = 
\left[\Theta, \Theta_1, \mY, \mZ, \Phi, \Pi \right] \exp{(\sigma t)}$
into equation~(\ref{eq:mini_gov_pert}) and derive
\begin{align}\label{eq:linker_normal}
 \lp 8\alpha+4 \rp\sigma \Theta \cos \bstheta + \sigma \Theta_1 \cos \bstheta + 
2\lp 3 -3 \alpha c + \cos 2\bstheta \rp \mY + 2 \mZ \sin2\bstheta  
&=0,\nonumber \\
 \lp 8\alpha+4 \rp\sigma \Theta \sin \bstheta+ \sigma \Theta_1 \sin \bstheta + 
2\lp 3 -3 
\alpha c - \cos 2\bstheta \rp \mZ + 2 \mY \sin2\bstheta  &=0,\nonumber \\
\lp 3\alpha+5/2 \rp \barmu \sigma \Theta + \barmu \sigma \Theta_1 + 
3\barmu \mY \cos\bstheta + 3\barmu \mZ \sin \bstheta - 24c \Theta_1 & 
=0,\nonumber \\
\lp 24 \alpha^2 + 21\alpha + 11/2\rp \barmu \sigma \Theta + 3\lp \alpha + 1/2 
\rp \barmu \sigma \Theta_1 + (24\alpha + 9)\barmu \lp \mY \cos\bstheta + \mZ 
\sin\bstheta  \rp \nonumber \\
+ 24c \Theta_1 - 24c \bareta \sigma \Theta + 24c\bareta \kappa \barE^2 \Theta + 
24 c \barE \lp\Phi \cos \bstheta - \Pi \sin \bstheta \rp &=0,\nonumber \\
\lp \sigma + \kappa \rp \Phi + \kappa^2 \bareta \barE \Theta \cos \bstheta &=0, 
\nonumber \\
\lp \sigma + \kappa \rp \Pi - \kappa^2 \bareta \barE \Theta \sin \bstheta &=0.
\end{align} 
Setting the determinant of the operator matrix of equation 
(\ref{eq:linker_normal}) to zero, we 
find the non-zero solutions of the complex growth rate $\sigma$
 \begin{dmath}\label{eq:sigma_linker}
  \sigma_{\pm} = \left\{\pm 
\sqrt{\Sigma} -6912 
\alpha  c^3 \bar{\eta }+288 c^2 \left\{ \alpha  \bar{\mu } \left[\kappa  
\bar{\eta 
}+24 \alpha  (\alpha +1)+8\right]-\alpha  \kappa  \barE^2 
\bar{\eta } \bar{\mu }+32 \bar{\eta }\right\}-3 c \bar{\mu } \left\{\alpha  
\left[ 12 
\alpha  (5 \alpha +3)+7\right] \kappa  \bar{\mu }-80 \kappa  \barE^2 
\bar{\eta }+80 \kappa  \bar{\eta }+256\right\}+4 \kappa  \bar{\mu }^2 
\right\}\\
/\left\{ 48 c \bar{\eta } (5-6 \alpha  
c)+\bar{\mu } \left[ 3 \alpha  (12 \alpha  (5 \alpha +3)+7\right] c-4)\right\},
 \end{dmath}
where
\begin{dmath}
 \scriptsize
 \Sigma= 
\left\{ 2304 c^2 \bar{\eta } (3 \alpha  c-4)-48 c \bar{\mu } \left[6 \alpha  c 
\left(\kappa  \bar{\eta }+24 \alpha  (\alpha +1)+8\right)-5 \kappa  \bar{\eta 
}-16\right]+\kappa  \bar{\mu }^2 \left[3 \alpha  (12 \alpha  (5 \alpha +3)+7) 
c-4\right]+48 
c \kappa  \barE^2 \bar{\eta } \bar{\mu } (6 \alpha  
c-5)\right\}^2-3072 c \kappa  \bar{\mu } \left\{48 c \bar{\eta } (6 \alpha  
c-5)-3 
\alpha   \left[ 12 \alpha  (5 \alpha +3)+7\right] c \bar{\mu }+4 \bar{\mu 
}\right\} \left\{3 c 
\bar{\eta} (4-3 \alpha  c)+\bar{\mu } \left[3 \alpha  (3 \alpha  (\alpha 
+1)+1) 
c-1\right]+3 c \barE^2 \bar{\eta } (3 \alpha  c-4)\right\}.
\end{dmath}
We plot $ \sigr$ and $\sigi$ in 
comparison with their numerical counterparts in figure~\ref{fig:results-linker}.

\section*{Acknowledgments}
We thank Drs. E. Han, L. Li, Y. Man and F. Yang, and Professors F. Gallaire, E. 
Nazockdast, O. S. Pak, B. Rallabandi  and Y. N. Young for
useful discussions. Prof. T. G\"{o}tz is acknowledged for sharing with us 
his PhD thesis. We thank the anonymous referees for their insightful comments. 
L.Z. thanks the Swedish Research Council for a VR International Postdoc Grant 
(2015-06334). We thank the NSF for 
support
via the Princeton University Material Research Science and Engineering Center 
(DMR-1420541). The computer time was
provided by SNIC (Swedish National Infrastructure for Computing).

\end{document}

%% file: symdef.inc
\def \be {\mathbf{e}}

\def \br {\mathbf{r}}

\def \taumw {\tau_{\mathrm{MW}}}
\def \taus  {\tau_{\mathrm{s}}}
\def \taupart  {\tau_{\mathrm{p}}}

\def \nP {P^{\ast}}
\def \sp {\sigma_{\mathrm{p}}}
\def \ss {\sigma_{\mathrm{s}}}
\def \ep {\epsilon_{\mathrm{p}}}
\def \es {\epsilon_{\mathrm{s}}}

\def \bP {\mathbfcal{P}}

\def \bE {\mathbf{E}}

\def \bT {\boldsymbol{\Gamma}}
\def \bF {\mathbf{F}}
\def \bI {\mathbf{I}}
\def \bf {\mathbf{f}}

\def \bU {\mathbf{U}}
\def \bUinf {\bU^{\infty}}

\def \nt {t^{\ast}}

\def \barmu {\bar{\mu}}

\def \barmucrione {\barmu_1^{\mathrm{cri}}}
\def \barmucritwo {\barmu_2^{\mathrm{cri}}}
\def \mbarmucrizero {\mbarmu_0^{\mathrm{cri}}}
\def \mbarmucrione {\mbarmu_1^{\mathrm{cri}}}
\def \mbarmucritwo {\mbarmu_2^{\mathrm{cri}}}

\def \bOmega {\boldsymbol{\Omega}}

\def \lp {\left(}
\def \rp {\right)}

\def \bx {\mathbf{x}}

\def \bxp {\mathbf{x}_{p}}

\def \d {\mathrm{d}}

\def \nE {E^{\ast}}
\def \bareta {\bar{\eta}}

\def \esl {\epsilon_{\mathrm{sld}}}

\def \muopt {\bar{\mu}^{\mathrm{opt}}}
\def \mUopt {\bar{\mathcal{U}}^{\mathrm{opt}}}
\def \mU {\bar{\mathcal{U}}}

\def \nonOmegalpk {\bar{\Omega}^{\mathrm{lpk}}}
\def \nonOmegaamp {\nonOmega^{\mathrm{mag}}}

\def \barEcrione {\bar{E}_{1}^{\mathrm{cri}}}
\def \barEcritwo {\bar{E}_{2}^{\mathrm{cri}}}

\def \mP {\mathcal{P}}
\def \mPinf {\mathcal{P}^{\infty}}

\def \bPinf {\bP^{\infty}}
\def \bPtt {\bP^{\mathrm{total}}}